\documentclass{article}

\usepackage{arxiv}

\usepackage[utf8]{inputenc} 
\usepackage[T1]{fontenc}    
\usepackage{hyperref}       
\usepackage{url}            
\usepackage{booktabs}       
\usepackage{amsfonts}       

\usepackage{enumitem}
\usepackage{tikz}
\usetikzlibrary{shapes, arrows, decorations.pathreplacing}
\usepackage{pgfplots}
\usepackage{subfig}
\usepackage{amsmath}
\usepackage[misc,geometry]{ifsym}

\tikzstyle{decision} = [diamond, draw, aspect=2, text width=5em, align=center, inner sep=0pt]
\tikzstyle{prop_decision} = [diamond, draw=red, aspect=2, text width=5em, align=center, inner sep=0pt]
\tikzstyle{block} = [rectangle, draw, text width=11em, text badly centered, minimum height=1em]
\tikzstyle{prop_block} = [rectangle, draw=red, text width=11em, text badly centered, minimum height=1em]
\tikzstyle{sblock} = [rectangle, draw, text width=3em, text badly centered, minimum height=1em]
\tikzstyle{mblock} = [rectangle, draw, text width=6.5em, text badly centered, minimum height=1em]
\tikzstyle{io} = [trapezium, trapezium left angle=70, trapezium right angle=110, text width=7.5em, minimum height=1cm, text centered, draw=black]
\tikzstyle{line} = [draw, -latex']

\definecolor{mycolor}{RGB}{128,128,128}

\usetikzlibrary{calc}
\tikzset{
	jump/.style={
		to path={
			let \p1=(\tikztostart),\p2=(\tikztotarget),\n1={atan2(\y2-\y1,\x2-\x1)} in
			(\tikztostart) -- ($($(\tikztostart)!#1!(\tikztotarget)$)!0.15cm!(\tikztostart)$)
			arc[start angle=\n1-180,end angle=\n1,radius=0.15cm] -- (\tikztotarget)}
	},
	jump/.default={0.5}
}

\definecolor{babyblue}{rgb}{0.54,0.81,0.94}
\definecolor{mypink}{rgb}{0.858, 0.188, 0.478}

\pgfplotsset{compat=1.10}

\title{Accelerating Matrix Factorization by Dynamic Pruning for Fast Recommendation}

\newif\ifuniqueAffiliation
\uniqueAffiliationtrue

\ifuniqueAffiliation 
\author{
	Yining Wu \\
	School of Computer Engineering and Science,\\
	Shanghai University,\\
	Shanghai, 200444, China\\
	\texttt{wuyining@shu.edu.cn} \\
	\And
	\href{https://orcid.org/0009-0000-9321-8380}{\includegraphics[scale=0.06]{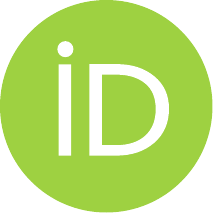}\hspace{1mm}Shengyu Duan (\Letter)} \\
	School of Computer Engineering and Science,\\
	Shanghai University,\\
	Shanghai, 200444, China\\
	\texttt{sduan@shu.edu.cn} \\
	\And
	Gaole Sai \\
	College of Integrated Circuits and Optoelectronic Chips,\\
	Shenzhen Technology University,\\
	Shenzhen, 518118, China\\
	\texttt{saigaole@sztu.edu.cn} \\	
	\And
	Chenhong Cao \\
	School of Computer Engineering and Science,\\
	Shanghai University,\\
	Shanghai, 200444, China\\
	\texttt{caoch@shu.edu.cn} \\
	\And
	Guobing Zou \\
	School of Computer Engineering and Science,\\
	Shanghai University,\\
	Shanghai, 200444, China\\
	\texttt{gbzou@shu.edu.cn} \\
}
\else
\usepackage{authblk}

\newbox{\orcid}\sbox{\orcid}{\includegraphics[scale=0.06]{orcid.pdf}} 
\author[1]{%
	\href{https://orcid.org/0000-0000-0000-0000}{\usebox{\orcid}\hspace{1mm}Yining Wu\thanks{\texttt{hippo@cs.cranberry-lemon.edu}}}%
}
\author[1,2]{%
	\href{https://orcid.org/0000-0000-0000-0000}{\usebox{\orcid}\hspace{1mm}Elias D.~Striatum\thanks{\texttt{stariate@ee.mount-sheikh.edu}}}%
}
\affil[1]{Department of Computer Science, Cranberry-Lemon University, Pittsburgh, PA 15213}
\affil[2]{Department of Electrical Engineering, Mount-Sheikh University, Santa Narimana, Levand}
\fi

\hypersetup{
pdftitle={A template for the arxiv style},
pdfsubject={q-bio.NC, q-bio.QM},
pdfauthor={David S.~Hippocampus, Elias D.~Striatum},
pdfkeywords={First keyword, Second keyword, More},
}

\begin{document}
\maketitle

\begin{abstract}
Matrix factorization (MF) is a widely used collaborative filtering (CF) algorithm for recommendation systems (RSs), due to its high prediction accuracy, great flexibility and high efficiency in big data processing.
However, with the dramatically increased number of users/items in current RSs, the computational complexity for training a MF model largely increases.
Many existing works have accelerated MF, by either putting in additional computational resources or utilizing parallel systems, introducing a large cost.
In this paper, we propose algorithmic methods to accelerate MF, without inducing any additional computational resources.
In specific, we observe fine-grained structured sparsity in the decomposed feature matrices when considering a certain threshold.
The fine-grained structured sparsity causes a large amount of unnecessary operations during both matrix multiplication and latent factor update, increasing the computational time of the MF training process.
Based on the observation, we firstly propose to rearrange the feature matrices based on joint sparsity, which potentially makes a latent vector with a smaller index more dense than that with a larger index. The feature matrix rearrangement is given to limit the error caused by the later performed pruning process.
We then propose to prune the insignificant latent factors by an early stopping process during both matrix multiplication and latent factor update. The pruning process is dynamically performed according to the sparsity of the latent factors for different users/items, to accelerate the process.
The experiments show that our method can achieve 1.2-1.65 speedups, with up to 20.08\% error increase, compared with the conventional MF training process. We also prove the proposed methods are applicable considering different hyperparameters including optimizer, optimization strategy and initialization method.
\end{abstract}


\section{Introduction}
In the present era of big data, the growing challenge of information overload emerges the need of information filtering systems to deal with massive data. 
Recommendation systems (RSs) are widely applied information filtering systems, which provide personalized suggestions of items ($e.g.,$ products, content or services) for certain users, among an overwhelming number of choices. 
However, the number of users and items involved in a RS is rapidly increased.
For instance, the commercial mobile application store, Google Play, has over one billion active users and over one million applications, and the number of applications has increased by approximately 1.8 million since 2013 \cite{r1}.
Training such at-scale RSs is evolving to require sufficiently large hardware resources and computational time.
It is necessary to efficiently accelerate the training process of RSs, with a minimal cost of extra resources. 

Typical filtering strategies of RSs include content-based, knowledge-based and collaborative filtering (CF) methods. 
Content-based methods have the problems of over-specialization and the difficulties to generate item-attribute information, as it is often based on certain business scenarios and manual extraction. Knowledge-based methods need a clear definition of explicit recommendation knowledge, introducing a great cost caused by knowledge acquisition, 
and the quality of its recommendations largely depends on the utilized knowledge base and user's feedback \cite{r2,r3,r4,r5}. 
CF is considered as the most commonly used method, due to its simplicity for implementation and high efficiency \cite{r6, r7}. 

CF-based RSs can be classified into two categories: neighborhood-based (also known as memory- or heuristic-based) methods and model-based methods \cite{r8}. 
Neighborhood-based methods leverage the similarity among multiple users or multiple items to make recommendations \cite{r9}.
Nevertheless, Neighborhood-based methods are less efficient in the situation with a large number of users and items, because such methods require a huge similarity matrix, the generation of which takes a large amount of time \cite{r10}.
Model-based CF typically uses machine learning (ML) algorithms and users' historical preferences or ratings to build a model and make recommendations. 
There are various methods to build the models, such as matrix factorization (MF) \cite{r11}, deep neural networks \cite{r12}, Latent Semantic Analysis \cite{r13}, Support Vector Machines \cite{r14} and Bayesian Clustering \cite{r15}. 
Compared with other methods, matrix factorization has the advantages of high prediction accuracy, great flexibility and high efficiency in big data processing \cite{r11,r16}.

In the MF method, all existing ratings are stored in a rating matrix, which is normally very sparse, as each user usually interacts with only a small subset in the overall item universe. 
MF decomposes the rating matrix by considering it as the inner product of two low-rank matrices, a user-feature matrix and an item-feature matrix, to recover the missing values of the rating matrix, in latent dimension.
Singular Value Decomposition (SVD) is one of the MF algorithms, used in RSs \cite{r11}.
However, the traditional SVD requires not only the decomposed matrices to be orthogonal, but also the rating matrix to be pre-filled, making the prediction accuracy largely affected by human factors \cite{r16,r18}. 
Some advanced algorithms have been proposed to address the problems of the conventional SVD, such as FunkSVD \cite{r19}, BiasSVD \cite{r20}, SVD++ \cite{r20}.

The computational complexity for training a MF model largely increases with more users and items \cite{ying2018graph}.
To improve the training efficiency of MF-based RSs, many existing works have proposed to parallelize the training process by Graphic Processing Units or distributed computer systems \cite{r24,r32}. However, in this way, a large investment in terms of hardware resources and power is required to further accelerate the process, and thus the cost is high.
In addition, we observe the decomposed user-/item-feature matrix exhibits fine-grained structured sparsity, as we will demonstrate in Section \ref{sec:sparsity}. This fine-grained structured sparsity causes a large amount of unnecessary computations to be involved in the training process of MF, which cannot be avoided if the process is parallelized. 

In this paper, we propose methods to dynamically prune the less significant features, to largely reduce the amount of unnecessary computations, while provide similar recommendations as the original MF method, without any additional hardware resources. The proposed methods are applicable to be operated on any computational platforms. The contributions of this paper are as follows:
\begin{enumerate}[label=\arabic*)]
	\item Given a certain threshold, we observe the presence of fine-grained structured sparsity in the decomposed matrices produced by MF, and the overall trend of sparsity for all latent vectors generally remains during the entire training process.
	\item We propose to rearrange the decomposed matrices based on joint sparsity as a pre-pruning process, to minimize the error caused by pruning.
	\item We propose a pruning approach, which is dynamically performed during the training process based on the sparsity of a certain latent dimension, to accelerate the process considering fine-grained structured sparsity.
\end{enumerate}

The rest of this paper is organized as follows. Section \ref{sec:background} introduces the typical MF algorithms and the commonly used optimization strategies. 
In Section \ref{sec:preliminary}, we investigate the overall process of MF-based RSs and demonstrate the fine-grained structured sparsity in the decomposed feature matrices.
Based on the fine-grained structured sparsity, we propose our pruning strategy, in Section \ref{sec:methodology}. 
In Section \ref{sec:results}, we provide experimental results, where it can be seen that our approach can realize 1.16-2.3 speedups with up to 20.08\% error increase. 
Finally, Section \ref{sec:conclusions} summarizes this paper.

\section{Background} \label{sec:background}
\subsection{MF Algorithms}


The traditional SVD decomposes a rating matrix by three matrices, where the inner product of the three matrices is considered to approximate the rating matrix, as follows:
\begin{equation}
\begin{aligned}
\boldsymbol{A}_{m\times n} &= \boldsymbol{U}_{m\times m} \textstyle\boldmath{\sum}_{m\times n}\boldsymbol{V}_{n\times n}\\
&\approx \boldsymbol{U}_{m\times k} \textstyle\boldmath{\sum}_{k\times k}\boldsymbol{V}_{k\times n}
\label{eq:SVD}
\end{aligned}
\end{equation}
where $\boldsymbol{A}_{m\times n}$ is a matrix with $m$ rows and $n$ columns, $\boldsymbol{U}_{m\times m}$ and $\boldsymbol{V}_{n\times n}$ are the left and right singular matrices, respectively, which are both orthogonal matrices, 
$\boldmath{\sum}_{m,n}$ is a diagonal matrix, where the diagonal elements are the singular values, arranged in descending order. 
In most cases, sum of the top a few singular values of $\boldmath{\sum}_{m\times n}$ is very close to the sum of all singular values,
so the singular values of the top $k$ $(k << m, n)$ are often used to approximate the matrix $\boldsymbol{A}_{m\times n}$ ($i.e.$, truncated SVD), to improve the computational efficiency and reduce the storage space \cite{r18}.

The traditional or truncated SVD requires no missing elements in the rating matrix.
However, a practical user-item rating matrix is very sparse as one user may only interact with a very small subset of all items.
A pre-filling process is therefore needed, but it may induce great data noise.
In addition, the decomposed matrices $\boldsymbol{U}_{m\times k}$ and $\boldsymbol{V}_{k\times n}$ are required to be orthogonal, and such a constraint further increases the prediction error. 

To further overcome the limitations of truncated SVD, FunkSVD directly decomposes the original sparse matrix into two dense matrices using only obtained ratings, by a training process, as shown in Equation (\ref{eq:funksvd}):
\begin{equation}
\boldsymbol{R}_{m\times n} \approx \boldsymbol{P}_{m\times k}\boldsymbol{Q}_{k\times n} \label{eq:funksvd}
\end{equation}
where $\boldsymbol{R}_{m\times n}$ represents the rating matrix, and the value of $m$ and $n$ depends on the number of users and items, respectively. 
$\boldsymbol{P}_{m\times k}$ and $\boldsymbol{Q}_{k\times n}$ are user-feature matrix and item-feature matrix, respectively.
FunkSVD projections the preferences of users and the attributes of items to a $k$-dimension latent feature space, and minimizes the error between the ratings generated by matrix multiplication and the actual ratings by gradient descent.
The objective function is as follows:
\begin{equation}
\min\limits_{\boldsymbol{P}_{m\times k},\boldsymbol{Q}_{k\times n}}\sum_{(u,i)\in \boldsymbol{\Omega}}{[(r_{u,i}-\boldsymbol{p}_u \boldsymbol{q}_i)^2 + \lambda(\|\boldsymbol{p}_u\|^2+\|\boldsymbol{q}_i\|^2)]}
\label{eq:obj-function}
\end{equation}
where $\boldsymbol{\Omega}$ denotes the index set of the actual ratings in $\boldsymbol{R}_{m\times n}$, $r_{u,i}$ ($r_{u,i }\in \boldsymbol{R}_{m\times n}$) is the actual rating given by the $u$-th user for the $i$-th item, $\boldsymbol{p}_u$ and $\boldsymbol{q}_i$ are a row and a column of $\boldsymbol{P}_{m\times k}$ and $\boldsymbol{Q}_{k\times n}$, respectively, representing the feature vectors of the $u$-th user and $i$-th item, respectively, and $\lambda$ is a regularization coefficient to avoid over-fitting.

Some improvements have been made based on FunkSVD, such as BiasSVD \cite{r20} and SVD++ \cite{r20}, to provide more accurate recommendations.
BiasSVD includes user bias, item bias and overall score of the training data, considering the subjective bias of each user and the quality of each item. 
SVD++ further includes some parameters to reveal implicit feedback and user's attribute information.

It is worth noting, in the rest of this work, we specifically investigate the training process of FunkSVD and accelerate FunkSVD by using our approaches. However, the proposed methods are also applicable for BiasSVD and SVD++, as they have the same training process as FunkSVD.

\subsection{Optimization Strategies in MF}
In order to solve the non-convex optimization problem given in Equation (\ref{eq:obj-function}), gradient descent is typically used to iteratively update the feature matrices. The process is also known as latent factor update. Stochastic Gradient Descent (SGD) is currently one of the most widely used gradient descent methods \cite{r21}.
Literally, SGD does not use all samples to calculate the gradient, but randomly selects a rating at a time to update the corresponding parameters. For each rating, $r_{u,i}$ ($r_{u,i }\in \boldsymbol{R}_{m\times n}$), $\boldsymbol{p}_u$ and $\boldsymbol{q}_i$ are updated as follows:
\begin{align}
e_{u,i} &= r_{u,i} - \boldsymbol{p}_u \boldsymbol{q}_i \label{eq:Sg-update} \\
\boldsymbol{p}_u &= \boldsymbol{p}_u + \alpha[e_{u,i}\boldsymbol{q}_i - \lambda \boldsymbol{p}_u] \label{eq:Sg-update-2} \\
\boldsymbol{q}_i &= \boldsymbol{q}_i + \alpha[e_{u,i}\boldsymbol{p}_u - \lambda \boldsymbol{q}_i] \label{eq:Sg-update-3}
\end{align}
where $\alpha$ is the learning rate, which needs to be adjusted based on the given applications, $e_{u,i}$ suggests the error between the predicted rating and the actual one given by the $u$-th user for the $i$-th item.
The above step is iteratively performed until appropriate $\boldsymbol{P}_{m\times k}$ and $\boldsymbol{Q}_{k\times n}$ are obtained for prediction, making the final recommendations.


The learning rate of a basic SGD approach is fixed, so it is crucial to set a reasonable learning rate. 
A learning rate which is too large or too small leads to divergence or slow convergence, respectively.
Optimizers providing adaptive learning rates are often necessary as an improvement to SGD.
These optimizers include Adagrad \cite{r26}, AdaDelta \cite{r27}, and Adam \cite{r28}.
Adagrad uses a different learning rate for each parameter and the learning rates decay according to the accumulated gradient, introducing large learning rates at the beginning of the training process and thus fast convergence.
AdaDelta is an extension of Adagrad that uses exponentially weighted average to retain only the past gradients of a certain window, leading to fast convergence once it nears the extremes.
What's more, with the addition of an exponentially weighted average of the update parameters, it eliminates the necessity to set a global learning rate, which significantly affects the overall prediction results.
Adam combines momentum and Adagrad, alleviating the problem caused by gradient oscillations.
The above mentioned three optimizers have been practically utilized in many MF-based RSs with high accuracy \cite{r29,r30,r31,chen2020accelerating}.

\section{Preliminaries} \label{sec:preliminary}
\subsection{Overall Process of MF-based RSs}

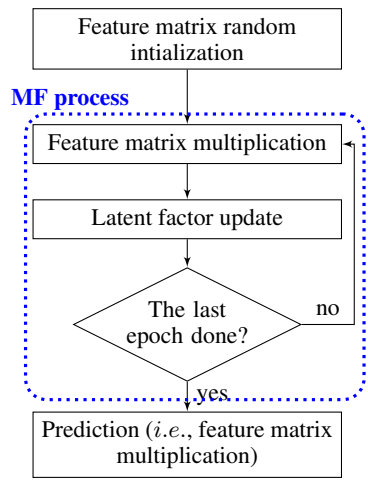
\begin{figure}[!htb]
	\centering
	\begin{tikzpicture}[node distance = 1.5cm, auto, font=\small]
	
	\node [block] (init) {Feature matrix random intialization};
	\node [block, below of=init, yshift=0.1cm] (mult) {Feature matrix multiplication};
	\node [block, below of=mult, yshift=0.5cm] (update) {Latent factor update};
	\node [decision, below of=update, yshift=0.1cm] (done) {The last epoch done?};
	\node [block, below of=done, yshift=-0.1cm] (pred) {Prediction ($i.e.$, feature matrix multiplication)}; 
	
	\path [line] (init) -- (mult);
	\path [line] (mult) -- (update);
	\path [line] (update) -- (done);
	\path [line] (done.south) -- node [midway]{yes}(pred.north);
	\path [line] (done.east) -- ++ (0.7,0) node[midway]{no} |- (mult.east);
	
	\draw (0.1,-2.9) node[draw, minimum width=4.5cm,minimum height=3.8cm, dotted, blue, very thick, rounded corners=8pt] (MF) {}
	(MF) node[above=2.1cm,left=0.75cm] {\textcolor{blue}{\textbf{MF process}}};
	
	\end{tikzpicture}
	\caption{Overall process of MF-based RSs.}
	\label{fig:Flowchart-LibMF}
\end{figure}

Fig. \ref{fig:Flowchart-LibMF} demonstrates a standard training process when using MF for RSs. An initialization stage is firstly required, which initializes the user-/item-feature matrix by filling randomized values.
During the MF process, inner product of the user-feature matrix and the item-feature matrix is produced, to generate all predicted ratings.
The errors are computed according to the actual ratings and the predicted ones, according to Equation (\ref{eq:Sg-update}), which are used to update all latent factors of the feature matrices, as given in Equations (\ref{eq:Sg-update-2}) and (\ref{eq:Sg-update-3}).
The above steps of MF-process involve a large number of matrix multiplications, determined by the number of users and items, and are iteratively performed over many epochs. The MF process is therefore time-consuming.
Recommendations can be made after the MF process is completed, where the ratings of all non-interacted items are predicted by the inner product of the feature matrices.
The above process is applied in many open-source libraries for MF and RSs, such as LibMF \cite{r33}, SVDFeature \cite{r34} and Surprise \cite{r35}.

We conduct a preliminary experiment using LibMF, to investigate the proportion of time spent on MF-process stage, by giving different number of epochs.
We perform MF-based recommendations on four datasets, MovieLens 100K \cite{ml}, Amazon Appliances \cite{appliances}, Book-Crossings \cite{bookCrossing} and Jester \cite{jester}, of which the detailed information is provided in Table \ref{tab:Datasets}.

\begin{table}[!htb]
	\centering
	\caption{Experimental datasets}
	\label{tab:Datasets}
	\setlength{\tabcolsep}{1mm}{
		\begin{tabular}{c|ccccc}
			\hline
			Dataset& \# users & \# items & \# ratings (training)& \# ratings (testing) & Rating scale\\
			\hline
			MovieLens 100K& 943& 1682& 90570& 9430& 1-5\\
			Appliances& 30252& 515650& 482221& 120556& 1-5\\
			Book-Crossings& 105284& 340554& 919823& 229956& 0-10 \\
			Jester& 73418& 100& 3308968& 827242& -10.0-10.0\\
			\hline
	\end{tabular}}
\end{table}

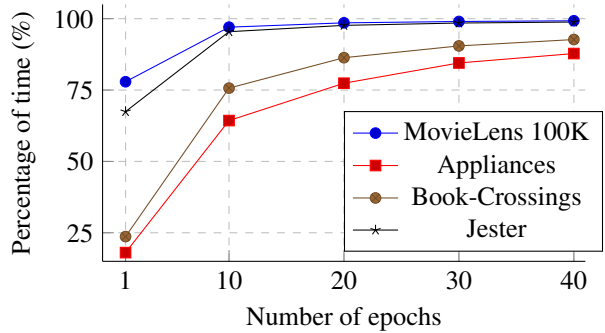
\begin{figure}[!htb]	
	\centering
	\begin{tikzpicture}[font=\normalsize, scale=1]	
	\begin{axis}[
	width=8cm,		
	height=5cm,		
	axis x line*=bottom, 
	axis y line*=left,
	ymax=105,
	ymin=15,
	xmin=-1,
	xmax=41,
	grid=major,
	grid style={dashed,gray!50},
	ylabel = {Percentage of time (\%)},
	xlabel = {Number of epochs},
	xticklabels={1,10,20,30,40},
	xtick={1,10,20,30,40},
	yticklabels={25,50,75,100},
	ytick={25,50,75,100},
	name=main plot,
	legend style={at={(axis cs:20,43)},anchor=west},
	clip=false
	]
	\addplot coordinates {(1,0.779045571200666*100) (10,0.9702964294375258*100) (20,0.9853937472237669*100) (30,0.9901654410208846*100) (40,0.9926812621144444*100)};
	\addlegendentry{MovieLens 100K}
	\addplot coordinates {(1,0.18053165041116848*100) (10,0.6430518098951847*100) (20,0.7739412867515109*100) (30,0.8449317064736829*100) (40, 0.8778431412905172*100)};
	\addlegendentry{Appliances}
	\addplot coordinates {(1,0.23671879600575413*100) (10,0.756773841027674*100) (20,0.863284749291819*100) (30,0.9046093823476262*100) (40, 0.9268897591268408*100)};
	\addlegendentry{Book-Crossings}	
	\addplot coordinates {(1,0.6745307481643906*100) (10,0.9543747384750416*100) (20,0.9765695076771932*100) (30,0.9847531159503902*100) (40, 0.9882285565465118*100)};
	\addlegendentry{Jester}			
	\end{axis}
	\end{tikzpicture}
	\caption{Proportion of time for MF in the overall process.}
	\label{fig:T-perc}
\end{figure}

Figure \ref{fig:T-perc} shows that the proportion of time spent in the MF-process stage gradually increases as the number of epochs increases.
When the number of epochs is greater than 10, 64.31\%-99.26\% of the overall time is taken to perform the MF process, while the training of a practical RS often requires hundreds of epochs, causing an even greater proportion of the total time to perform MF.
According to Figure \ref{fig:Flowchart-LibMF}, a MF epoch is generally consist of matrix multiplication and latent factor update, so we propose to accelerate both steps by dynamic pruning, as will be described in Section \ref{sec:methodology}, leading to a fast training process for MF-based RSs.

\subsection{Fine-grained Structured Sparsity in MF-based RSs} \label{sec:sparsity}

Sparse matrices are commonly found in the applications of ML algorithms including MF \cite{r36}. 
In CF-based RSs, sparsity often indicates the density of the user-item rating matrix, revealing the interactions of users for different items. 
The number of users and items is extremely large, and each user has only rated a limited number of items, resulting in a very sparse rating matrix.
Data sparsity of the rating matrix is inevitable.
In this paper, we don't discuss the sparsity of the rating matrix, instead we observe there exist a large number of insignificant features, which are close to zero, in the decomposed user-/item-feature matrices, indicating certain latent vectors of the feature matrices may contain more insignificant elements, as we will explain it as follows.   

\begin{figure}[!htb]
	\centering
	\begin{tikzpicture}[font=\footnotesize, scale=1]	
	\draw[draw=mycolor, fill=mycolor] (0,0) rectangle ++(0.2,0.2) node[pos=0.5] (zero) {};
	\draw (zero) node [right=0.05cm] {insignificant};	
	\draw[draw=mycolor] (2,0) rectangle ++(0.2,0.2) node[pos=0.5] (not-zero) {};
	\draw (not-zero) node [right=0.05cm] {significant};	
	\end{tikzpicture}
	
	\vspace{-0.4cm}
	\subfloat[]{\includegraphics[width=6cm]{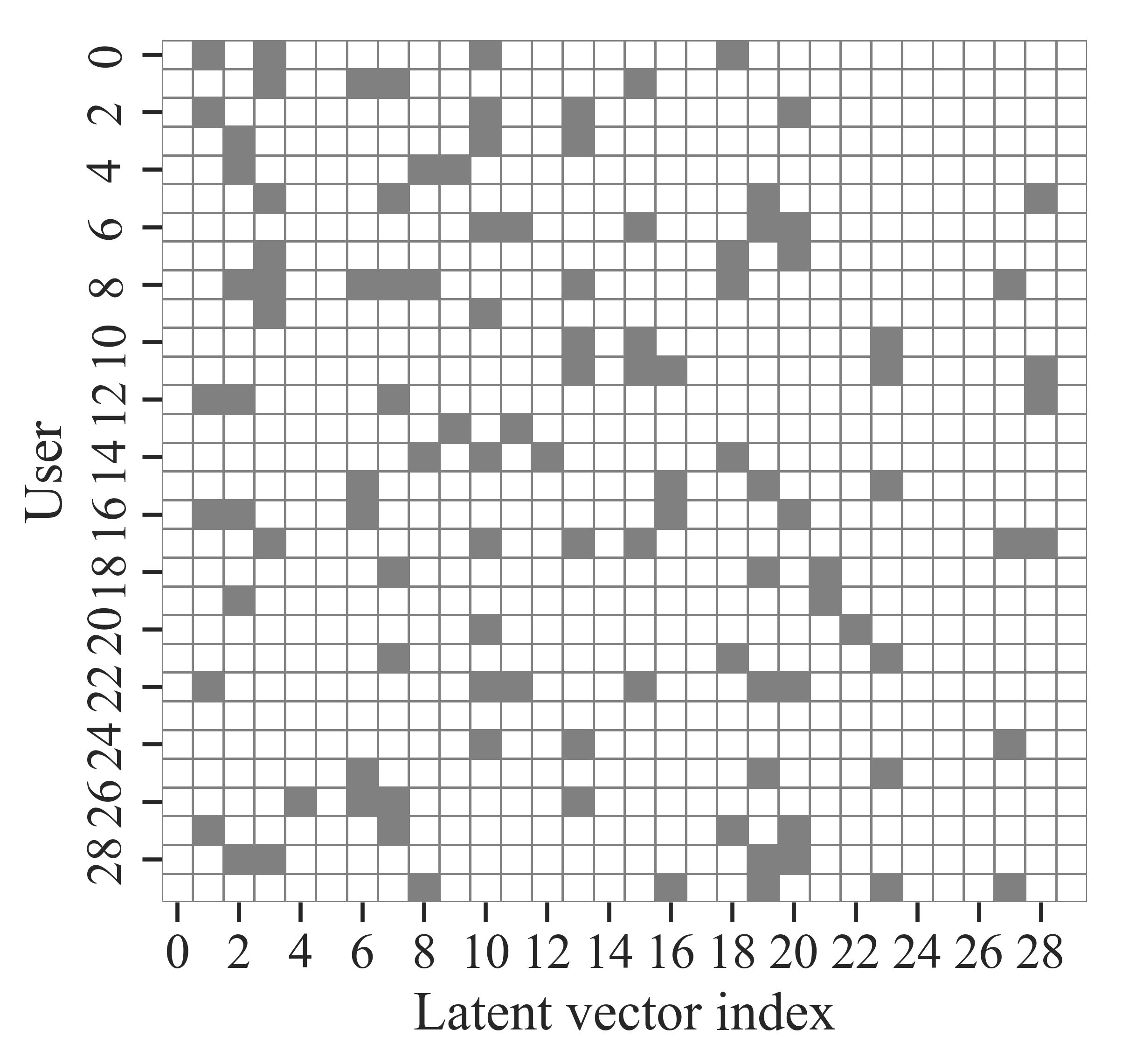}}
	\hspace{0.8cm}
	\subfloat[]{\includegraphics[width=6cm]{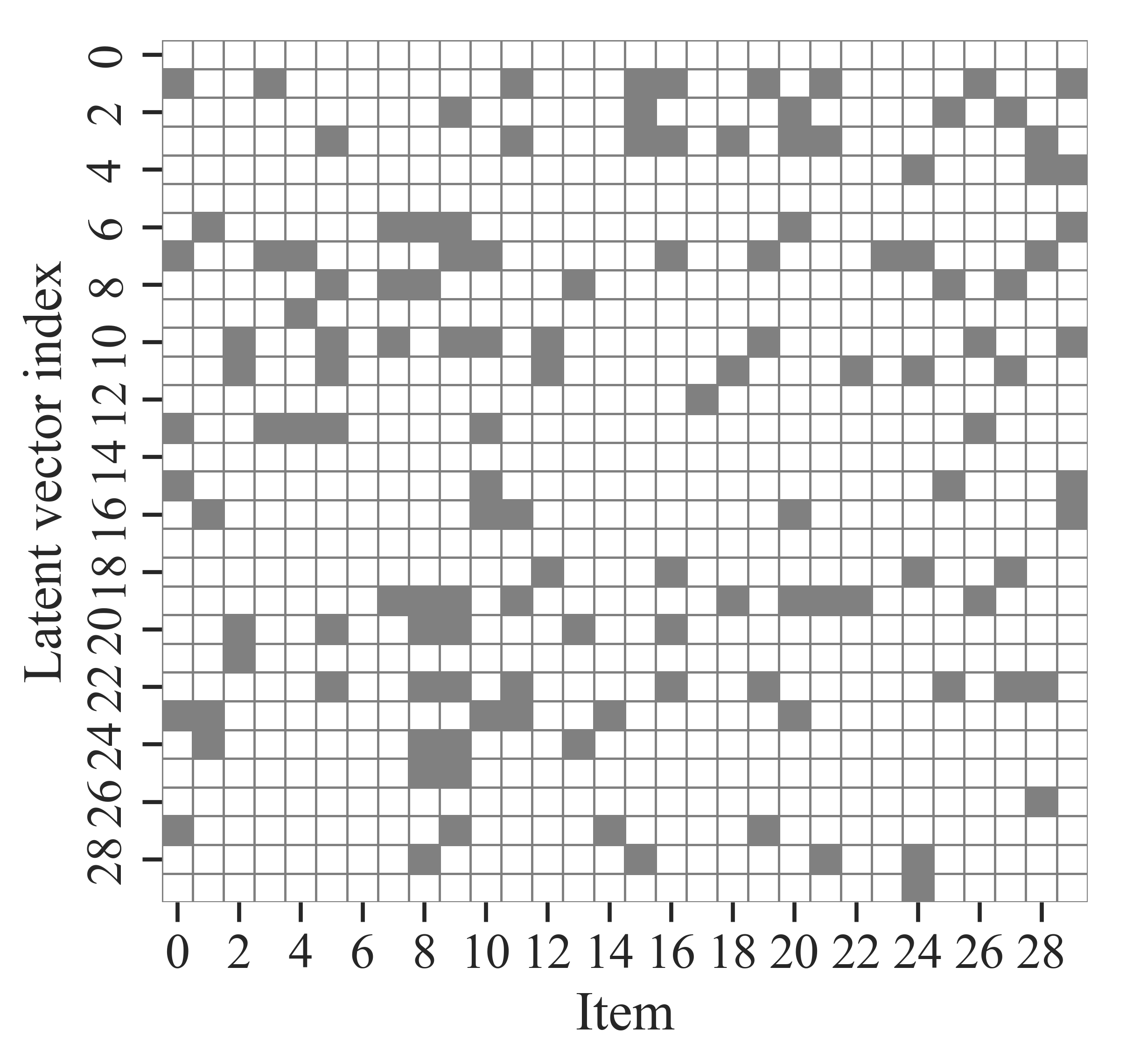}}
	\caption{Fine-grained structured sparsity of (a). $\boldsymbol{P}_{m\times k}$, and (b). $\boldsymbol{Q}_{k\times n}$.}
	\label{fig:Fine-grained}
\end{figure}
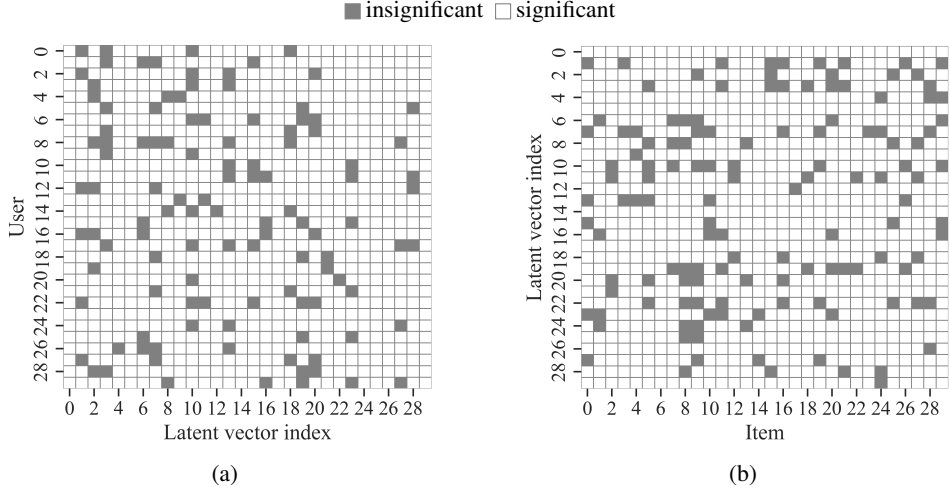

We firstly perform MF on the dataset, MovieLens 100K. We assume the rating matrix as $\boldsymbol{R}_{m\times n}$, which is decomposed into the user-feature matrix $\boldsymbol{P}_{m\times k}$ and the item-feature matrix $\boldsymbol{Q}_{k\times n}$. We denote $m$, $n$ and $k$ as the number of users, items and latent vectors, respectively. Figure \ref{fig:Fine-grained} demonstrates parts of $\boldsymbol{P}_{m\times k}$ and $\boldsymbol{Q}_{k\times n}$ by setting $k$ as 30, after 25 training epochs. 
The insignificant elements are defined by the elements with their absolute values less than a threshold, 0.06, in such a case.
Fine-grained structured sparsity can be observed according to Figure \ref{fig:Fine-grained}: insignificant elements are irregularly allocated in $\boldsymbol{P}_{m\times k}$ and $\boldsymbol{Q}_{k\times n}$,
and some latent vectors exhibit higher sparsity than the rest, in both matrices.
These partially sparse dimensions cannot be deleted through feature selection, as it may cause great errors.
However, as the insignificant elements are very close to zero, they generally do not contribute to the overall results when calculating the inner product of $\boldsymbol{P}_{m\times k}$ and $\boldsymbol{Q}_{k\times n}$, causing unnecessary computations and additional computational time.

\begin{figure}[!htb]	
	\centering
	\begin{tikzpicture}[font=\normalsize, scale=1]	
	\begin{axis}[
	width=8cm,		
	height=5cm,		
	axis x line*=bottom, 
	axis y line*=left,
	ymax=0.8,
	ymin=0.5,
	xmin=5,
	xmax=95,
	grid=major,
	grid style={dashed,gray!50},
	ylabel = {MAE},
	xlabel = {Number of latent vectors},
	xticklabels={10,30,50,70,90},
	xtick={10,30,50,70,90},
	name=main plot,
	legend style={at={(axis cs:45,0.75)},anchor=west},
	legend columns=2, 
	clip=false
	]
	\addplot coordinates {(10,0.7645500000000001) (30,0.737004) (50,0.6331149999999999) (70,0.615769) (90,0.562851)};
	\addlegendentry{Uid 21}
	\addplot coordinates {(10,0.530481) (30,0.564305) (50,0.6494740000000001) (70,0.6357779999999998) (90,0.665657)};
	\addlegendentry{Uid 24}
	\addplot coordinates {(10,0.6532589999999999) (30,0.6837310000000001) (50,0.657709) (70,0.6340669999999999) (90,0.6765270000000001)};
	\addlegendentry{Uid 33}
	\addplot coordinates {(10,0.6114710000000001) (30,0.5530379999999999) (50,0.571056) (70,0.556305) (90,0.5888239999999999)};
	\addlegendentry{Uid 6}					
	\end{axis}
	\end{tikzpicture}
	\caption{MAE between predicted and actual ratings for different number of latent vectors $k$.}
	\label{fig:U-MAE}
\end{figure}
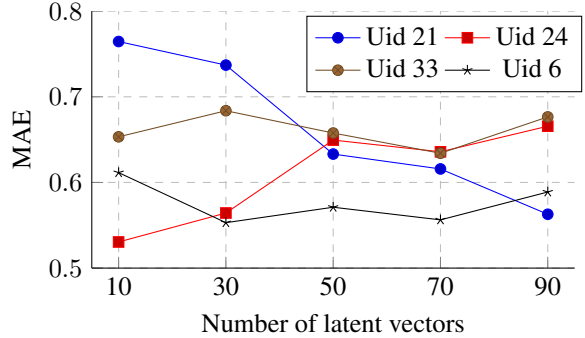

It might be argued that the fine-grained structured sparsity of $\boldsymbol{P}_{m\times k}$ and $\boldsymbol{Q}_{k\times n}$ can be eliminated by properly selecting the number of latent vectors, $k$.
By definition, $\boldsymbol{P}_{m\times k}$ and $\boldsymbol{Q}_{k\times n}$ are more dense by reducing the value of $k$, as the latent factors tend to be larger.
On the other hand, as the value of $k$ increases, elements of $\boldsymbol{P}_{m\times k}$ and $\boldsymbol{Q}_{k\times n}$ become smaller, causing greater sparsity.
Nevertheless, the suitable value of $k$ is often determined by empirical study in practice, to ensure the greatest prediction accuracy.
In Figure \ref{fig:U-MAE}, we provide the mean absolute errors (MAEs) between the predicted and actual ratings given by certain users for all items of MovieLens 100K, when $k$ increases.
MAE also indicates the degree of fit of the model when predicting the ratings for certain users: lower MAE suggests the predicted ratings are more accurate, and thus the model is better fitted.
As can be noticed from Figure \ref{fig:U-MAE}, the prediction accuracy for some users ($e.g.$, Uid 21) is significantly increased, as $k$ increases,
but for other users, the prediction accuracy generally remains unchanged ($e.g.$, Uid 6 and 33), or tends to decrease ($e.g.$, Uid 24).
This can be expected, because the preferences of some users might be more predictable than the rest, so a model with a certain value of $k$ tends to be better fitted when predicting the ratings of these users.
Increasing the value of $k$ can theoretically improve the prediction accuracy for those users, whose intentions are less predictable, but it may introduce more insignificant latent factors in the latent factor dimensions of other users or items, causing fine-grained structured sparsity in $\boldsymbol{P}_{m\times k}$ and $\boldsymbol{Q}_{k\times n}$.
As the predictability of different users or items is certainly different, we conclude this fine-grained structured sparsity in $\boldsymbol{P}_{m\times k}$ and $\boldsymbol{Q}_{k\times n}$ is unavoidable.

\section{Dynamic Pruning for Accelerated MF} \label{sec:methodology}
\subsection{Overall Procedure} \label{sec:overall}	
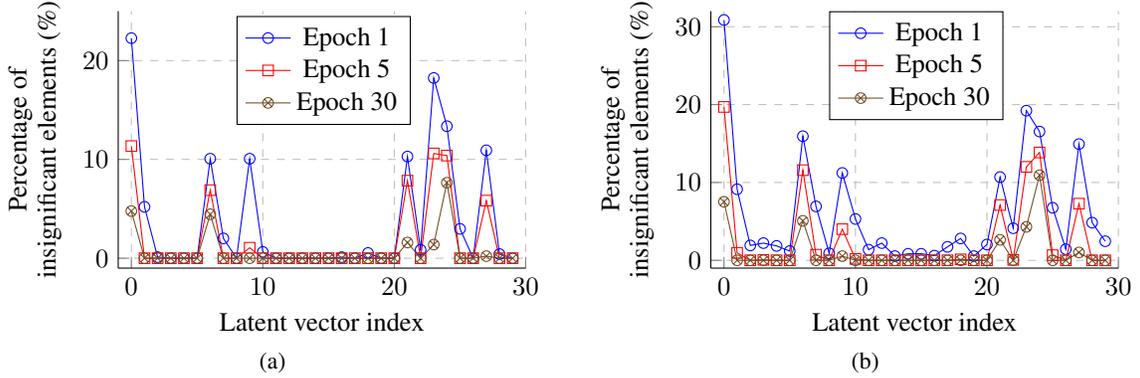
\begin{figure}[!htb]
	\centering
	\subfloat[]{
		\begin{tikzpicture}[font=\normalsize, scale=1]	
		\begin{axis}[
		width=7cm,		
		height=5cm,		
		axis x line*=bottom, 
		axis y line*=left,
		ymax=25,
		ymin=-1,
		xmin=-1,
		xmax=30,
		grid=major,
		grid style={dashed,gray!50},
		ylabel = {\begin{tabular}{c} Percentage of \\ insignificant elements (\%) \end{tabular}},
		xlabel = {Latent vector index},
		name=main plot,
		legend style={at={(axis cs:8,19)},anchor=west},
		clip=false
		]
		\addplot [color=blue, mark=o] table [y expr={\thisrow{p_t1}*100}, x=i] {data/sparsity.txt};
		\addlegendentry{Epoch 1}
		\addplot [color=red, mark=square] table [y expr={\thisrow{p_t5}*100}, x=i] {data/sparsity.txt};
		\addlegendentry{Epoch 5}
		\addplot [color=brown!60!black, mark=otimes] table [y expr={\thisrow{p_t30}*100}, x=i] {data/sparsity.txt};
		\addlegendentry{Epoch 30}						
		\end{axis}
		\end{tikzpicture}
	}
	\hspace{0.3cm}
	\subfloat[]{
		\begin{tikzpicture}[font=\normalsize, scale=1]	
		\begin{axis}[
		width=7cm,		
		height=5cm,		
		axis x line*=bottom, 
		axis y line*=left,
		ymax=32,
		ymin=-1,
		xmin=-1,
		xmax=30,
		grid=major,
		grid style={dashed,gray!50},
		ylabel = {\begin{tabular}{c} Percentage of \\ insignificant elements (\%) \end{tabular}},
		xlabel = {Latent vector index},
		name=main plot,
		legend style={at={(axis cs:8,25)},anchor=west},
		clip=false
		]
		\addplot [color=blue, mark=o] table [y expr={\thisrow{q_t1}*100}, x=i] {data/sparsity.txt};
		\addlegendentry{Epoch 1}
		\addplot [color=red, mark=square] table [y expr={\thisrow{q_t5}*100}, x=i] {data/sparsity.txt};
		\addlegendentry{Epoch 5}
		\addplot [color=brown!60!black, mark=otimes] table [y expr={\thisrow{q_t30}*100}, x=i] {data/sparsity.txt};
		\addlegendentry{Epoch 30}						
		\end{axis}
		\end{tikzpicture}
	}
	\caption{Sparsity of all latent factors after different numbers of epoch. (a) $\boldsymbol{P}_{m\times k}$. (b) $\boldsymbol{Q}_{k\times n}$.}
	\label{fig:Sparsity-k}
\end{figure}

During the training process of a MF model, the sparsity of each latent vector of matrices $\boldsymbol{P}_{m\times k}$ and $\boldsymbol{Q}_{k\times n}$ may vary with the number of epochs. 
Figure  \ref{fig:Sparsity-k} shows the sparsity of all latent vectors in both feature matrices, after 10, 20 and 30 training epochs using MovieLens 100K, where $k$ is set to be 30, 
and a feature is considered to be insignificant if its absolute value is less than 0.06. 
It can be seen that the sparsity of each latent vector tends to decrease with more training epochs, as the percentage of the insignificant elements generally decreases in both $\boldsymbol{P}_{m\times k}$ and $\boldsymbol{Q}_{k\times n}$.
However, the overall trends of the sparsity of all latent vectors generally hold: certain latent vectors exhibit higher sparsity than the rest, during the entire training process.
In this paper, we exploit the above observation to accelerate the training process of MF, as follows.

\begin{figure}[!htb]
	\centering
	\begin{tikzpicture}[node distance = 1.5cm, auto, font=\small]
	
	\node [block] (init) {Feature matrix random intialization};
	\node [block, below of=init, yshift=-0.5cm] (mult) {Feature matrix multiplication};
	\node [block, below of=mult, yshift=0.5cm] (update) {Latent factor update};
	\node [decision, below of=update, yshift=0.1cm] (done) {The last epoch done?};
	\node [block, below of=done, yshift=-0.1cm] (pred) {Prediction ($i.e.$, feature matrix multiplication)}; 
	
	\path [line] (mult) -- (update);
	\path [line] (update) -- (done);
	\path [line] (done.south) -- node [midway]{yes}(pred.north);
	
	\draw (0.1,-3.5) node[draw, minimum width=4.5cm,minimum height=3.8cm, dotted, blue, very thick, rounded corners=8pt] (MF) {}
	(MF) node[above=2.1cm,left=0.75cm] {\textcolor{blue}{\textbf{MF process}}};
	
	\node [prop_decision, below of=init, yshift=0.1cm, xshift=6cm] (first) {The first epoch?};
	\node [prop_decision, below of=first, yshift=-0.3cm] (second) {The second epoch?};
	\node [prop_block, below of=second, yshift=0.1cm] (thres) {Threshold determination};	
	\node [prop_block, below of=thres, yshift=0.5cm] (rearrange) {Feature matrix rearrangement};
	\node [prop_block, below of=rearrange, yshift=0.5cm] (prune) {Latent factor pruning};
	
	\path [line] (init.south) -- ++(0,-0.075) -| (first.north);
	\path [line] (first.south) -- node [midway]{no}(second.north);
	\path [line] (second.south) -- node [midway]{yes}(thres);
	\path [line] (thres) -- (rearrange);
	\path [line] (rearrange) -- (prune);
	\path [line] (second.east) -- ++ (0.8,0) node[midway]{no} |- (prune.east);
	\draw (first.west) node[above=0.15cm, left=0cm] {yes};
	\path [line] (done.east) -- ++ (0.7,0) node[midway]{no} -- ++ (0,3.915) -| (first.north);
	\draw (first.west) -- ++ (-0.8,0) |- (prune.west);
	\draw (first.west) -- ++ (-2,0) to[jump] ++ (-1,0);
	\node[left of = first, xshift=-2.6cm] (a) {};
	\path [line] (a) -| (mult.north);
	
	\draw (6,-3.65) node[draw, minimum width=4.9cm,minimum height=6.7cm, dotted, red!80!black, very thick, rounded corners=8pt] (prop) {}
	(prop) node[above=3.6cm,left=0cm] {\textcolor{red}{\textbf{Proposed methods}}};
	
	\end{tikzpicture}
	\caption{Overall procedure of the proposed methods.}
	\label{fig:Flowchart-method}
\end{figure}
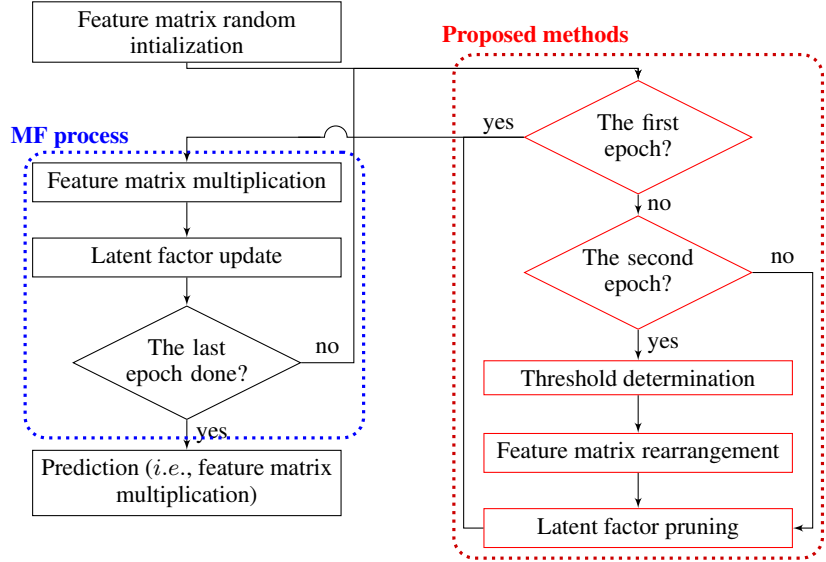

The overall procedure of the proposed methods is given, in Figure \ref{fig:Flowchart-method}.	
As can be seen, the proposed methods can be implemented with a standard MF training process.
Considering the fine-grained structured sparsity of the feature matrices, we firstly propose to rearrange $\boldsymbol{P}_{m\times k}$ and $\boldsymbol{Q}_{k\times n}$ into more coarse-grained structures, based on the sparsity of both matrices and a given threshold value for the later pruning process.
The above step only needs to be performed once after the first training epoch, because the overall trends of the sparsity of all latent vectors would hold during the entire training process, as demonstrated in Figure \ref{fig:Sparsity-k}.
Therefore, the more coarse-grained structured sparsity, introduced by the matrix rearrangement, can be preserved during the entire training process.
We then propose to perform coarse-grained pruning for the insignificant features, to accelerate both matrix multiplications and latent factor update.
As the sparsity of each latent vector in $\boldsymbol{P}_{m\times k}$ and $\boldsymbol{Q}_{k\times n}$ is different, the pruning process is dynamically facilitated when calculating with the latent factors of different users/items.
We provide the detailed descriptions of the proposed methods in Sections \ref{sec:threshold}-\ref{sec:pruning}.

\subsection{Determination of Threshold Value for Pruning} \label{sec:threshold}
A threshold value is required to determine significant/insignificant latent factors in $\boldsymbol{P}_{m\times k}$ and $\boldsymbol{Q}_{k\times n}$.
A higher threshold suggests a higher pruning rate, where more latent factors will be considered as insignificant and thus pruned later.
By definition, the higher the pruning rate is, the more computational time will be reduced to provide better acceleration, but the greater errors will be induced.
As the exact values of the latent factors may greatly vary among different applications, we propose to determine the threshold value based on a given pruning rate.

\begin{figure}[!htb]
	\centering
	\subfloat[]{\includegraphics[width=7cm]{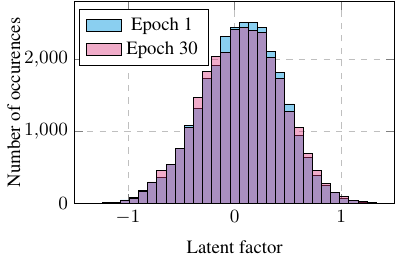}}
	\hspace{0.3cm}
	\subfloat[]{\includegraphics[width=7cm]{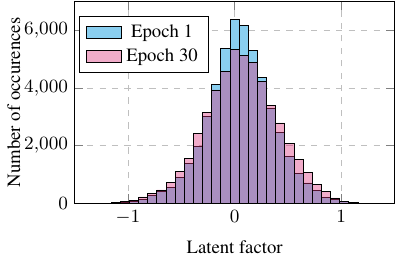}}
	
	\subfloat[]{\includegraphics[width=7cm]{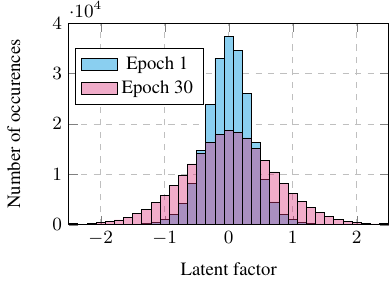}}
	\hspace{0.3cm}
	\subfloat[]{\includegraphics[width=7cm]{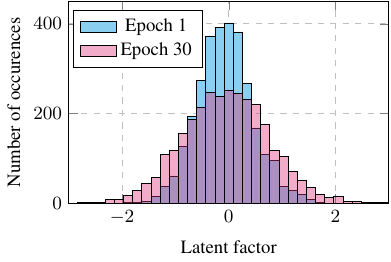}}
	\caption{Distributions of all latent factors, after training with 1 and 30 epochs. (a) MovieLens 100K, $\boldsymbol{P}_{m\times k}$. (b) MovieLens 100K, $\boldsymbol{Q}_{k\times n}$. (c) Jester, $\boldsymbol{P}_{m\times k}$. (d) Jester, $\boldsymbol{Q}_{k\times n}$.}
	\label{fig:Distribution-T}
\end{figure}

In Figure \ref{fig:Distribution-T}, we demonstrate the distributions of all latent factors in both $\boldsymbol{P}_{m\times k}$ and $\boldsymbol{Q}_{k\times n}$, after training with 1 and 30 epochs on MovieLens 100K and Jester. In each case, the latent factors exhibit a normal-like distribution, which can be used to compute the threshold value for a certain pruning rate.

Here we assume that the given pruning rate is $p$, to determine a suitable threshold value $T$. 
We assume the feature matrices obey a normal distribution with the mean value and the standard deviation as $\mu$ and $\sigma$, respectively, which can be measured once the feature matrices are firstly produced just after one training epoch.
We define the cumulative distribution function of a standard normal distribution as $\phi(x)$.
In order to determine $T$, which makes $p$ around latent factors locating in the range between $-T$ and $T$ and thus being pruned after the first epoch, we provide Equation (\ref{eq:Constraint2}) as follows: 
\begin{gather}	
T = \sigma x + \mu \label{eq:Constraint2}	
\end{gather}
where $x$ is a parameter satisfying Equation (\ref{eq:Constraint1}), which can be determined by searching it in a standard normal table.
\begin{gather}	
\phi(x) - \phi(-x - 2\mu/\sigma) = p \label{eq:Constraint1}
\end{gather}

The detailed explanation and derivation of Equations (\ref{eq:Constraint2}) and (\ref{eq:Constraint1}) are provided, in the Appendix.

The process to determine $T$ is thereby as follows:
\begin{enumerate}[label=\arabic*)]
	\item Find $x$ that satisfies Equation (\ref{eq:Constraint1}) in a standard normal distribution;
	\item Statistically measure $\mu$ and $\sigma$ for all latent factors of the two feature matrices and calculate $T$, according to Equation (\ref{eq:Constraint2}).
\end{enumerate}

\begin{figure}[!htb]
	\centering
	\subfloat[]{
		\begin{tikzpicture}[font=\normalsize, scale=1]	
		\begin{axis}[
		width=7cm,		
		height=5cm,		
		axis x line*=bottom, 
		axis y line*=left,
		ymax=50,
		ymin=13,
		xmin=0,
		xmax=31,
		grid=major,
		grid style={dashed,gray!50},
		ylabel = {\begin{tabular}{c} Percentage of \\ insignificant elements (\%) \end{tabular}},
		xlabel = {Number of epochs},
		name=main plot,
		legend style={at={(axis cs:3,45)},anchor=west, font=\small},
		clip=false,
		legend columns=2
		]
		\addplot table [y expr={\thisrow{p_ml}*100}, x=i] {data/sparsity_epoch.txt};
		\addlegendentry{MoiveLens 100K}
		\addplot table [y expr={\thisrow{p_app}*100}, x=i] {data/sparsity_epoch.txt};
		\addlegendentry{Appliances}
		\addplot table [y expr={\thisrow{p_bx}*100}, x=i] {data/sparsity_epoch.txt};
		\addlegendentry{Book-Crossings}
		\addplot table [y expr={\thisrow{p_jes}*100}, x=i] {data/sparsity_epoch.txt};
		\addlegendentry{Jester}							
		\end{axis}
		\end{tikzpicture}
	}\hspace{0.3cm}
	\subfloat[]{
		\begin{tikzpicture}[font=\normalsize, scale=1]	
		\begin{axis}[
		width=7cm,		
		height=5cm,		
		axis x line*=bottom, 
		axis y line*=left,
		ymax=50,
		ymin=13,
		xmin=0,
		xmax=31,
		grid=major,
		grid style={dashed,gray!50},
		ylabel = {\begin{tabular}{c} Percentage of \\ insignificant elements (\%) \end{tabular}},
		xlabel = {Latent vector index},
		name=main plot,
		legend style={at={(axis cs:4,45)},anchor=west, font=\small},
		clip=false,
		legend columns=2
		]
		\addplot table [y expr={\thisrow{q_ml}*100}, x=i] {data/sparsity_epoch.txt};
		\addlegendentry{MoiveLens 100K}
		\addplot table [y expr={\thisrow{q_app}*100}, x=i] {data/sparsity_epoch.txt};
		\addlegendentry{Appliances}
		\addplot table [y expr={\thisrow{q_bx}*100}, x=i] {data/sparsity_epoch.txt};
		\addlegendentry{Book-Crossings}
		\addplot table [y expr={\thisrow{q_jes}*100}, x=i] {data/sparsity_epoch.txt};
		\addlegendentry{Jester}							
		\end{axis}
		\end{tikzpicture}
	}
	\caption{Sparsity after different training epochs. (a) $\boldsymbol{P}_{m\times k}$. (b) $\boldsymbol{Q}_{k\times n}$.}
	\label{fig:Sparsity-T}
\end{figure}
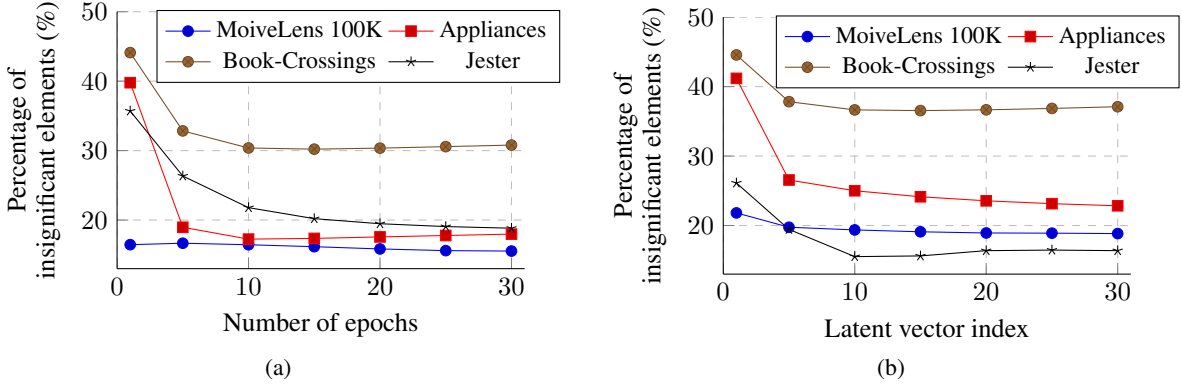

It is worth noting the threshold value is determined by using the above method only once, just after the first training epoch, rather than in each epoch,
because the overall distributions of all latent factors do not significantly change after the first epoch, as shown in Figure \ref{fig:Distribution-T}.
Figure \ref{fig:Sparsity-T} demonstrates the sparsity of both $\boldsymbol{P}_{m\times k}$ and $\boldsymbol{Q}_{k\times n}$ after different training epochs on different datasets. 
As can be seen, $\boldsymbol{P}_{m\times k}$ and $\boldsymbol{Q}_{k\times n}$ tend to be less sparse in each case, which is consistent as the results shown in Figure \ref{fig:Sparsity-k}.
Therefore, a fewer latent factors are considered to be insignificant and will be pruned later, once a threshold is determined after the first epoch.
We consider the threshold determined by $\boldsymbol{P}_{m\times k}$ and $\boldsymbol{Q}_{k\times n}$ after the first epoch as a pessimistic value, leading to the most insignificant latent factors to be pruned and thus the greatest reduction for computational time.
Although this may induce additional error, one can adjust the pruning rate to trade off between computational time and accuracy.

\subsection{Feature Matrix Rearrangement based on Joint Sparsity} \label{sec:rearrange}
Once the threshold values to determine the insignificant latent factors are properly selected, the sparsity of $\boldsymbol{P}_{m\times k}$ and $\boldsymbol{Q}_{k\times n}$ can be evaluated to rearrange both matrices. 
As the inner product of $\boldsymbol{P}_{m\times k}$ and $\boldsymbol{Q}_{k\times n}$ is computed during each training epoch, the unnecessary multiplications are determined considering the insignificant latent factors in both $\boldsymbol{P}_{m\times k}$ and $\boldsymbol{Q}_{k\times n}$.
An unnecessary multiplication indicates there is at least one latent factor of the two is insignificant.
Therefore, for certain threshold values, $T_p$ and $T_q$, for $\boldsymbol{P}_{m\times k}$ and $\boldsymbol{Q}_{k\times n}$, respectively, we give the definition of joint sparsity for the $k$-th latent vectors, denoted by  $JS_k$, in Equation (\ref{eq:Joint-sparsity}).   
\begin{align}
JS_k &= prob(\lvert \boldsymbol{P}_{\{1:m\},k}\lvert < T_p \cap \lvert \boldsymbol{Q}_{k,\{1:n\}}\lvert < T_q)  	\label{eq:Joint-sparsity} 	\\
&= prob(\lvert \boldsymbol{P}_{\{1:m\},k}\lvert < T_p) \times p(\lvert \boldsymbol{Q}_{k,\{1:n\}}\lvert < T_q) \label{eq:Joint-sparsity-2}
\end{align}
where $\boldsymbol{P}_{\{1:m\},k}$ and $\boldsymbol{Q}_{k,\{1:n\}}$ represent the $k$-th latent vectors of $\boldsymbol{P}_{m\times k}$ and $\boldsymbol{Q}_{k\times n}$, respectively, $prob(\lvert \boldsymbol{P}_{\{1:m\},k}\lvert < T_p)$ and $prob(\lvert \boldsymbol{Q}_{k,\{1:n\}}\lvert < T_q)$ are the probabilities that the factors of $\boldsymbol{P}_{\{1:m\},k}$ and $\boldsymbol{Q}_{k,\{1:n\}}$, respectively, are considered as insignificant. Here we assume $\boldsymbol{P}_{m\times k}$ and $\boldsymbol{Q}_{k\times n}$ are independent, so $JS_k$ can be computed using Equation (\ref{eq:Joint-sparsity-2}). 

We provide Algorithm 1 to rearrange $\boldsymbol{P}_{m\times k}$ and $\boldsymbol{Q}_{k\times n}$ based on $JS_k$, which potentially makes a column/row with a smaller index more dense than that with a larger index. By performing Algorithm 1, the rearranged $\boldsymbol{P}_{m\times k}$ and $\boldsymbol{Q}_{k\times n}$ satisfy Equation (\ref{eq:After-rearranged}).
\begin{equation}
\forall k_1, k_2 \in [1,k] \land k_1 < k_2: JS_{k_1} < JS_{k_2}
\label{eq:After-rearranged}
\end{equation}

\begin{table}[!htb]
	\begin{tabular*}{\textwidth}{@{\extracolsep{\fill}} l }
		\toprule
		\textbf{ALGORITHM 1:} Feature matrix rearrangement based on joint sparsity\\
		\midrule
		\textbf{Input:} $\boldsymbol{P}_{m\times k} = \{\boldsymbol{P}_{\{1:m\},1}, \boldsymbol{P}_{\{1:m\},2}, ..., \boldsymbol{P}_{\{1:m\},k}\}$, \\
		\ \ \hspace{2em} $\boldsymbol{Q}^\top_{k\times n} = \{\boldsymbol{Q}_{\{1:n\},1}, \boldsymbol{Q}_{\{1:n\},2}, ..., \boldsymbol{Q}_{\{1:n\},k}\}$, $T_p$, $T_q$ 
		\\
		\textbf{Output:} rearranged $\boldsymbol{P}_{m\times k}$ and $\boldsymbol{Q}_{k\times n}$\\
		1: \ \textbf{for} $i$ = \{1,...,$k$\} \textbf{do}\\
		2: \ \hspace{1em} calculate $JS_i$ according to Equation (\ref{eq:Joint-sparsity-2})\\
		3: \ \textbf{end for}\\
		4: \ \textbf{for} $i$ = \{1,...,$k$-1\} \textbf{do}\\
		5: \ \hspace{1em} \textbf{for} $j$ = \{2,...,$k$\} \textbf{do}\\
		6: \ \hspace{2em} \textbf{if} $JS_i < JS_j$\\
		7: \ \hspace{3em} swap $\boldsymbol{P}_{\{1:m\},i}$/$\boldsymbol{Q}_{\{1:n\},i}$ with $\boldsymbol{P}_{\{1:m\},j}$/$\boldsymbol{Q}_{\{1:n\},j}$ \\
		\ \ \hspace{4em} in $\boldsymbol{P}_{m\times k}$/$\boldsymbol{Q}_{k\times n}$\\
		8: \ \hspace{2em} \textbf{end if}\\
		9: \ \hspace{1em} \textbf{end for}\\
		10: \textbf{end for}\\
		\bottomrule
	\end{tabular*}
\end{table}

Algorithm 1 only needs to be executed once after the first epoch, because the overall trends of the sparsity of all latent vectors generally hold during the entire MF training process, as has been explained in Section \ref{sec:overall}.

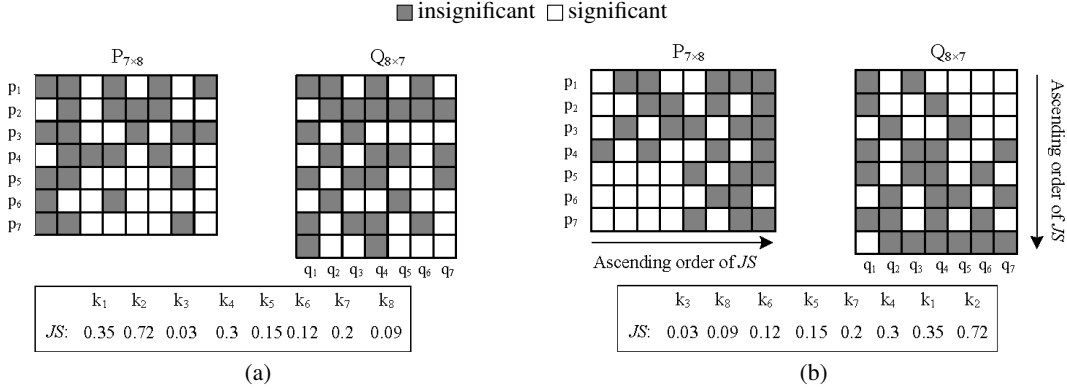
\begin{figure}[!htb]
	\centering
	\begin{tikzpicture}[font=\footnotesize, scale=1]	
	\draw[draw=black, fill=mycolor] (0,0) rectangle ++(0.2,0.2) node[pos=0.5] (zero) {};
	\draw (zero) node [right=0.05cm] {insignificant};	
	\draw[draw=black] (2,0) rectangle ++(0.2,0.2) node[pos=0.5] (not-zero) {};
	\draw (not-zero) node [right=0.05cm] {significant};	
	\end{tikzpicture}
	
	\vspace{-0.4cm}
	\subfloat[]{\includegraphics[width=7cm]{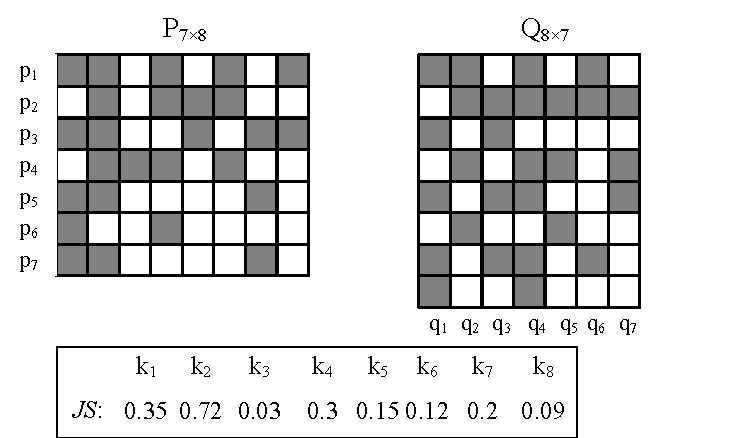}}\hspace{0.3cm}
	\subfloat[]{\includegraphics[width=7cm]{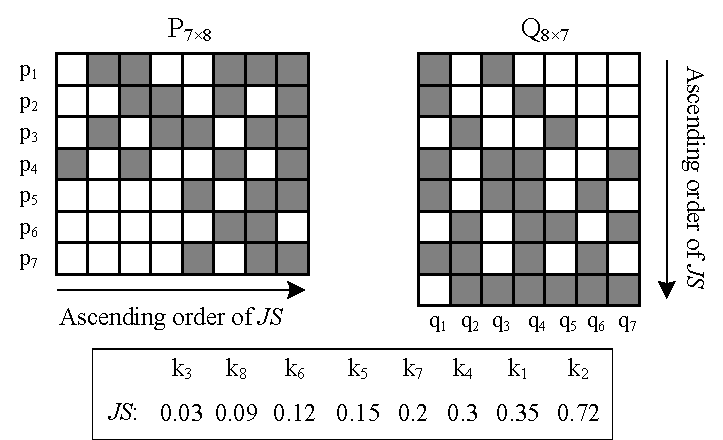}}
	\caption{An example of feature matrix rearrangement. (a) Original matrices. (b) Rearranged matrices.}
	\label{fig:Rearranged}
\end{figure}

We provide Figure \ref{fig:Rearranged} as an example to rearrange two feature matrices, $\boldsymbol{P}_{7\times 8}$ and $\boldsymbol{Q}_{8\times 7}$, by using Algorithm 1. As can be seen, the insignificant latent factors are originally irregularly allocated in  $\boldsymbol{P}_{7\times 8}$ and $\boldsymbol{Q}_{8\times 7}$. The joint sparsity for each latent vector is computed according to Equation (\ref{eq:Joint-sparsity-2}), and given in Figure \ref{fig:Rearranged} (a). 
By performing Algorithm 1, $\boldsymbol{P}_{7\times 8}$ and $\boldsymbol{Q}_{8\times 7}$ are rearranged to ensure the joint sparsity in an ascending order, as illustrated in Figure \ref{fig:Rearranged} (b).
It can be noticed from Figure \ref{fig:Rearranged} (b) that more significant latent factors appear in the first a few columns/rows of $\boldsymbol{P}_{7\times 8}$/$\boldsymbol{Q}_{8\times 7}$, after the rearrangement.       

\subsection{Dynamic Pruning for Fast Matrix Multiplication and Latent Factor Update} \label{sec:pruning}
Matrix multiplication of $\boldsymbol{P}_{m\times k}$ and $\boldsymbol{Q}_{k\times n}$ is conducted during each MF training epoch.
Dot product of each row and column of $\boldsymbol{P}_{m\times k}$ and $\boldsymbol{Q}_{k\times n}$ is involved during a standard matrix multiplication, which is typically performed starting from the elements with small indices.
After rearranging the feature matrices using Algorithm 1, the columns or rows with smaller indices, of $\boldsymbol{P}_{m\times k}$ or $\boldsymbol{Q}_{k\times n}$, respectively, tend to be less sparse, as has been explained in Section \ref{sec:rearrange}.
Therefore, for the multiplication of certain vectors, $\boldsymbol{p}_u$ ($\boldsymbol{p}_u\in \boldsymbol{P}_{m\times k}$) and $\boldsymbol{q}_i$ ($\boldsymbol{q}_i\in \boldsymbol{Q}_{k\times n}$), we can expect the multiplication of the significant latent factors is potentially performed earlier than that of the insignificant ones.
We thereby propose Algorithm 2 to mostly prune those insignificant latent factors, leading to fast matrix multiplication.  
According to Algorithm 2, for each row-by-column multiplication, latent factors of the row vector, $\boldsymbol{p}_u$, are multiplied by the corresponding ones of the column vector, $\boldsymbol{q}_i$, till the first insignificant latent factor appears, in either $\boldsymbol{p}_u$ or $\boldsymbol{q}_i$, leading to early stopping to reduce the computational time. 
We expect the latent factors that are pruned minimally contribute to the row-by-column multiplication result, because they are more likely to be insignificant in the rearranged feature matrices.
The amount of error caused by the early stopping process is thereby limited.  	
\begin{table}[!htb]
	\begin{tabular*}{\textwidth}{@{\extracolsep{\fill}} l }
		\toprule
		\textbf{ALGORITHM 2:} Fast and approximate matrix multiplication by early stopping\\
		\midrule
		\textbf{Input:} $\boldsymbol{p}_u$, $\boldsymbol{q}_i$, $T_p$, $T_q$, where $\boldsymbol{p}_u \in \boldsymbol{P}_{m\times k}$, $\boldsymbol{q}_i \in \boldsymbol{Q}_{k\times n}$,\\
		\hspace{3em} $u \in [1,m]$, $i \in [1,n]$\\
		\textbf{Output:} $\boldsymbol{p}_u . \boldsymbol{q}_i$, which is the predicted rating of the $i$-th user\\ 								\ \hspace{3em} for the $j$-th item\\
		1: \ $\boldsymbol{p}_u . \boldsymbol{q}_i = 0$\\
		2: \ \textbf{for} $t$ = \{1,...,$k$\} \textbf{do}\\
		3: \ \hspace{1em} \textbf{if} ($|p_{u,t}|<T_p$) or ($|q_{t,i}|<T_q$) \textbf{do} /* where the $p_{u,t}$\\
		\ \hspace{3em} and $q_{t,i}$ is the t-th element of $\boldsymbol{p}_u$ and $\boldsymbol{q}_i$, respectively*/\\
		4: \ \hspace{2em} \textbf{break}\\
		5: \ \hspace{1em} \textbf{else} \ $\boldsymbol{p}_u . \boldsymbol{q}_i = \boldsymbol{p}_u . \boldsymbol{q}_i + p_{u,t} . q_{t,i}$\\
		6: \ \hspace{1em} \textbf{end if}\\
		7: \ \textbf{end for}\\
		\bottomrule
	\end{tabular*}
\end{table}

Through Algorithm 2, the dot product of each row and column of $\boldsymbol{P}_{m\times k}$ and $\boldsymbol{Q}_{k\times n}$ can be approximately obtained, reducing the computational time to calculate $e_{u,i}$ of Equation (\ref{eq:Sg-update}).
$e_{u,i}$ is then used to update the feature vectors of certain users and items by gradient descent, as given in Equations (\ref{eq:Sg-update-2}) and (\ref{eq:Sg-update-3}).
We provide Algorithm 3 to mostly prune the insignificant latent factors from the update process.
According to Section \ref{sec:sparsity}, the optimal number of latent dimensions for each user and item are different.
When all feature vectors have the same size, it might cause overfitting for the feature vectors of some users and items.
Therefore, in Algorithm 3, we propose not to update the insignificant elements of the feature matrices, which not only reduces the overall computational time by eliminating the update process of some latent factors, but prevents overfitting to ensure high prediction accuracy.
\begin{table}[!htb]
	\begin{tabular*}{\textwidth}{@{\extracolsep{\fill}} l }
		\toprule
		\textbf{ALGORITHM 3:} Accelerated gradient descent by pruning \\
		\midrule
		\textbf{Input:} $\boldsymbol{p}_u$, $\boldsymbol{q}_i$, $T_p$, $T_q$, where $\boldsymbol{p}_u \in \boldsymbol{P}_{m\times k}$, $\boldsymbol{q}_i \in \boldsymbol{Q}_{k\times n}$,\\
		\hspace{3em} $u \in [1,m]$, $i \in [1,n]$\\
		\textbf{Output:} $\boldsymbol{p}_u$ and $\boldsymbol{q}_i$ that have been updated\\
		1: \ \textbf{for} $t$ = \{1,...,$k$\} \textbf{do}\\
		2: \ \hspace{1em} \textbf{if} ($|p_{u,t}|<T_p$) or ($|q_{t,i}|<T_q$) \textbf{do} /* where the $p_{u,t}$\\
		\ \hspace{3em} and $q_{t,i}$ is the t-th element of $\boldsymbol{p}_u$ and $\boldsymbol{q}_i$, respectively*/\\
		3: \ \hspace{2em} \textbf{break}\\
		4: \ \hspace{1em} \textbf{else} \ update $p_{u,t}$ and $q_{t,i}$ \\
		5: \ \hspace{1em} \textbf{end if}\\
		6: \ \textbf{end for}\\
		\bottomrule
	\end{tabular*}
\end{table}

\begin{figure}[!htb]
	\centering
	\begin{tikzpicture}[node distance = 1.5cm, auto, font=\small]
	
	\node [io] (in) {\begin{tabular}{c} Row vector $\boldsymbol{p}_u$, \\ Column vector $\boldsymbol{q}_i$, \\ $j$=1 \end{tabular}};
	\node [decision, below of=in, yshift=-1.3cm] (sign) {$p_u$ or $q_i$ insignificant?};
	\node [mblock, right of=sign, xshift=1.7cm] (rate) {Predicted rating += $p_u*q_i$};
	\node [decision, right of=rate, xshift=2cm] (done) {Vector multiplication done?};
	\node [sblock, above of=done, yshift=0.1cm] (j) {$j$ += 1};
	\node [sblock, below of=sign, yshift=0cm] (j2) {$j$=1};
	
	\path [line] (in.south) -- (sign.north);
	\path [line] (sign.east) -- node [midway]{no}(rate.west);
	\path [line] (rate.east) -- (done.west);
	\path [line] (done.north) -- node [midway]{no}(j.south);
	\path [line] (j.west) -| (sign.north);
	\path [line] (sign.south) -- node[midway]{yes}(j2.north);
	\path [line] (done.south) -- ++ (0,-0.2) node[midway]{yes} -| (j2.north);
	
	\node [decision, below of=j2, yshift=-1cm] (sign2) {$p_u$ or $q_i$ insignificant?};
	\node [mblock, right of=sign2, xshift=1.7cm] (update) {Update $p_u$ and $q_i$};
	\node [decision, right of=update, xshift=2cm] (done2) {All factor updated?};
	\node [sblock, above of=done2, yshift=0.1cm] (j3) {$j$ += 1};
	\node [mblock, below of=sign2, yshift=-0.1cm] (next) {Next row/column};
	
	\path [line] (j2.south) -- (sign2.north);
	\path [line] (sign2.east) -- node [midway]{no}(update.west);
	\path [line] (update.east) -- (done2.west);
	\path [line] (done2.north) -- node [midway]{no}(j3.south);
	\path [line] (j3.west) -| (sign2.north);
	\path [line] (sign2.south) -- node[midway]{yes}(next.north);
	\path [line] (done2.south) -- ++ (0,-0.4) node[midway]{yes} -| (next.north);
	
	\draw (3.5,-2.4) node[draw, minimum width=10cm,minimum height=3.1cm, dotted, red, very thick, rounded corners=8pt] (prune1) {}
	(prune1) node[above=1.8cm,right=0.5cm] {\textcolor{red}{\textbf{Matrix multiplication with pruning}}};
	
	\draw (3.5,-6.45) node[draw, minimum width=10cm,minimum height=3.2cm, dotted, red, very thick, rounded corners=8pt] (prune2) {}
	(prune2) node[above=1.8cm,right=0.5cm] {\textcolor{red}{\textbf{Latent factor update with pruning}}};
	
	\end{tikzpicture}
	\caption{Overall dynamic pruning process.}
	\label{fig:Pruning}
\end{figure}
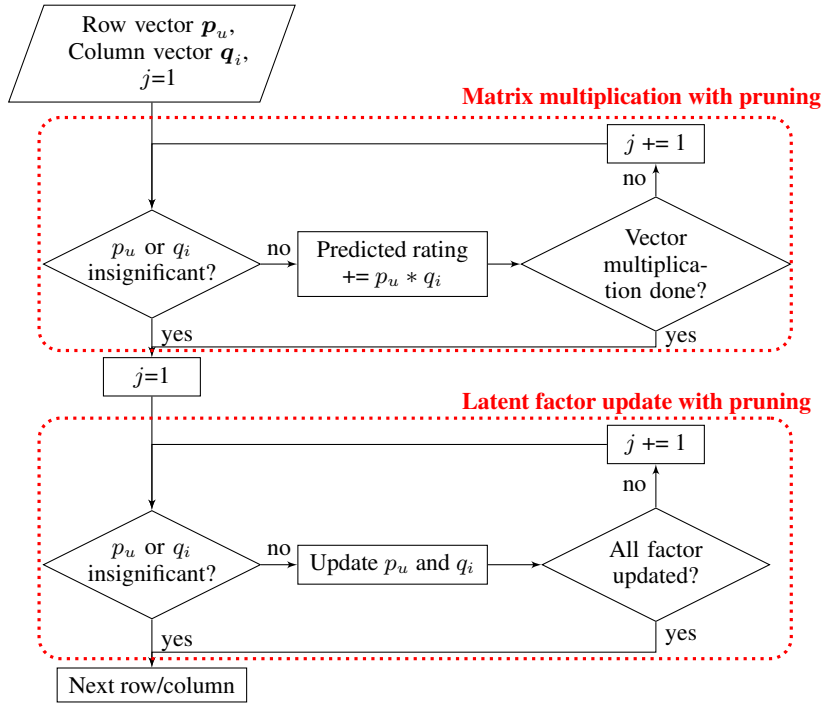

The overall pruning process is provided, in Figure \ref{fig:Pruning}. 
By definition, the previous feature matrix rearrangement ensures the latent factors that are pruned in both matrix multiplication and latent factor update generally insignificant to affect the overall results, reducing the amount of error increase induced by pruning.
The pruning process is performed for all existing ratings in each training epoch, except the first epoch, as the feature matrices are rearranged during the second epoch. 
As the sparsity of each user and item feature vectors of $\boldsymbol{P}_{m\times k}$ and $\boldsymbol{Q}_{k\times n}$, respectively, is different, and the sparsity keeps changed during all training epochs, the pruning process is dynamically performed, based on the actual sparsity of certain users or items and of certain epochs.

\section{Results}\label{sec:results}

\subsection{Experimental Setup}
We conduct experiments on the four datasets, including MovieLens 100K, Amazon Appliances, Book-Crossings, and Jester, of which the details are given in Table \ref{tab:Datasets}.
For each dataset, we randomly select 80\% and 20\% of all ratings as the training and test sets, respectively. 

Our experiments are conducted based on the open-source RS framework LibMF \cite{r33}, which implements FunkSVD as the MF algorithm and Adagrad as the optimizer to update the latent factors. 
Before the first epoch, we initialize all latent factors by a normal distribution.
We specifically investigate the impact of different hyperparameters from the above mentioned ones, to the performance of our proposed methods, in Section \ref{sec:impact}.
The experiments are performed on a machine with Intel i5-12500H 2.5GHz processors.
We openly provide all source code of this work at \href{https://github.com/Git-SmSun/DP-MF}{https://github.com/Git-SmSun/DP-MF}.

We evaluate the performance of the proposed methods by MAE, percentage MAE ($P_{MAE}$) and speedup, given in Equations (\ref{eq:mae})-(\ref{eq:speedup}), respectively.		
\begin{gather}
MAE = \frac{1}{N}\sum_{r_{u,i}\in \boldsymbol{R}}|r_{u,i}-r'_{u,i}| \label{eq:mae} \\
P_{MAE} = \frac{MAE_{acc}-MAE_{org}}{MAE_{org}} \times 100\% \label{eq:mae2}
\end{gather}
where $N$ is the number of actual ratings in the test set, $\boldsymbol{R}$ is the rating matrix, $r_{u,i}$ and $r'_{u,i}$ are the actual and predicted ratings, respectively, given by the $u$-th user for the $i$-th item, $MAE_{org}$ and $MAE_{acc}$ are the MAE produced by the conventional and accelerated training process, respectively.	
\begin{equation}
Speedup = \frac{t_{org}}{t_{acc}} \label{eq:speedup}
\end{equation}
where ${t_{org}}$ and ${t_{acc}}$ are the total runtime when not using and using the proposed methods for acceleration, respectively. The runtime is measured including the time of initialization, MF process and prediction, as introduced in Section \ref{sec:overall}.

\subsection{Speedup and Prediction Accuracy}

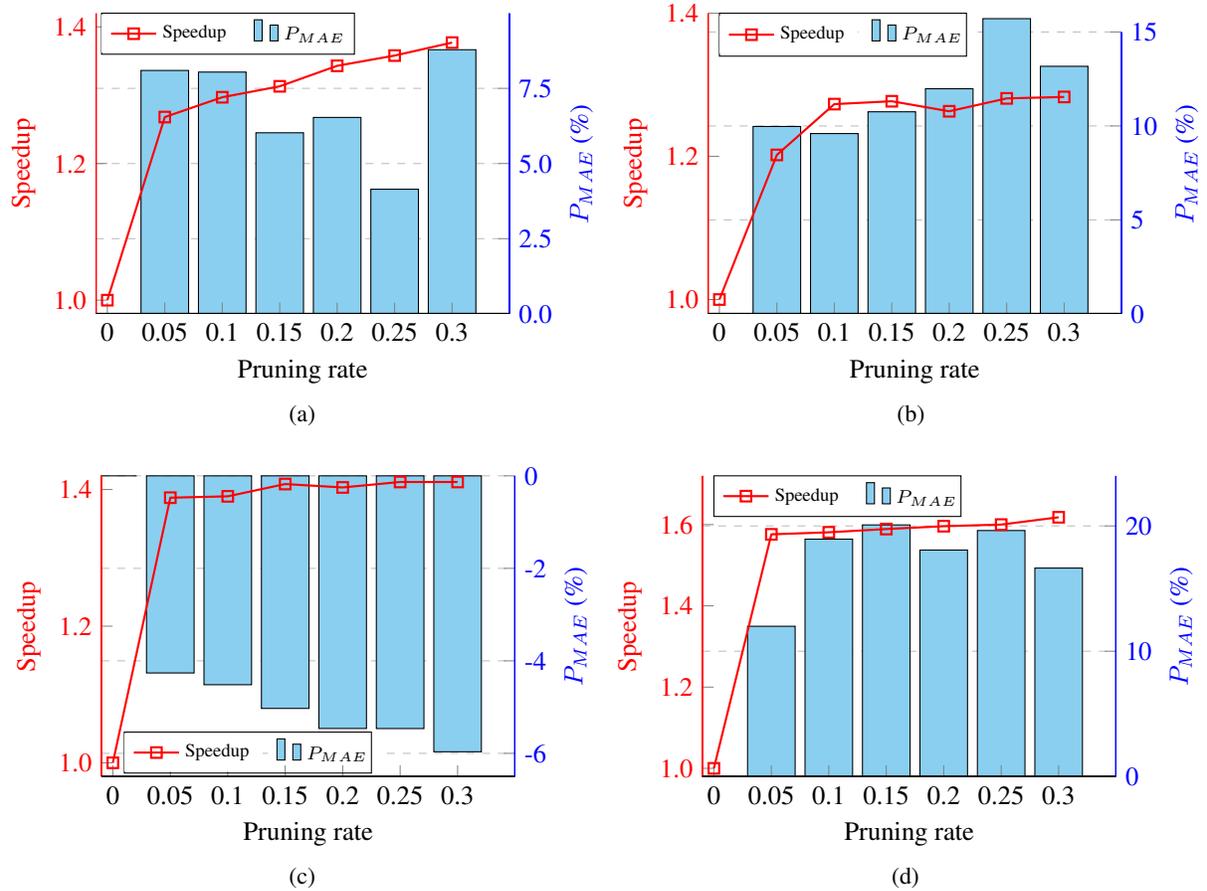
\begin{figure}[!htb]
	\centering
	\subfloat[]{
		\begin{tikzpicture}[scale=1]
		\pgfplotsset{
			scale only axis,
		}
		
		\begin{axis}[
		ybar,
		bar width=18pt,
		width=5.5cm,		
		height=4cm,		
		axis x line=none,
		axis y line*=right,
		axis line style={white},
		ymax=10,
		ymin=0,
		xmin=-0.01,
		xmax=0.35,
		ytick={0,2.5,5.0,7.5},
		yticklabels={\textcolor{blue}{0.0},\textcolor{blue}{2.5},\textcolor{blue}{5.0},\textcolor{blue}{7.5}},
		ylabel = {\textcolor{blue}{$P_{MAE}$ (\%)}},
		legend style={at={(axis cs:0.12,9.3)},anchor=west, font=\scriptsize},
		legend style={draw=none},
		grid=major,
		grid style={dashed,gray!50},
		]	
		\addplot[style={fill=babyblue,mark=none}] coordinates {(0,0.0) (0.05,8.095770157718478) (0.1,8.041165429808611) (0.15,6.022837467618407) (0.2,6.5329273914722945) (0.25,4.1452514724866925) (0.3,8.785172328496746)};
		\addlegendentry{$P_{MAE}$}	
		\draw [thick,blue] (axis cs:0.35,0) -- node[]{} (axis cs:0.35,10); 
		\end{axis}
		
		\begin{axis}[
		width=5.5cm,		
		height=4cm,		
		axis x line*=bottom, 
		axis y line*=left,
		axis line style={white},
		ymax=1.42,
		ymin=0.98,
		xmin=-0.01,
		xmax=0.35,
		ytick={1.0,1.2,1.4},
		yticklabels={\textcolor{red}{1.0},\textcolor{red}{1.2},\textcolor{red}{1.4}},
		xtick={0,0.05,0.1,0.15,0.2,0.25,0.3},
		xticklabels={0,0.05,0.1,0.15,0.2,0.25,0.3},
		ylabel = {\textcolor{red}{Speedup}},
		xlabel = {Pruning rate},
		legend style={at={(axis cs:-0.005,1.39)},anchor=west, font=\scriptsize},
		legend style={draw=none}]		
		\addplot [red, mark=square, thick] coordinates {(0,1) (0.05,1.268) (0.1,1.297) (0.15,1.313) (0.2,1.343) (0.25,1.358) (0.3,1.377)};
		\addlegendentry{Speedup}
		\draw [thick,red] (axis cs:-0.01,0.98) -- node[]{} (axis cs:-0.01,1.42);
		\draw [thick,black] (axis cs:-0.01,0.98) -- node[]{} (axis cs:0.35,0.98);
		\draw [draw=black] (axis cs: -0.005,1.36) rectangle (axis cs: 0.215, 1.42);  	
		\end{axis}
		
		\end{tikzpicture}
	}
	\subfloat[]{
		\begin{tikzpicture}[scale=1]
		\pgfplotsset{
			scale only axis,
		}
		
		\begin{axis}[
		ybar,
		bar width=18pt,
		width=5.5cm,		
		height=4cm,		
		axis x line=none,
		axis y line*=right,
		axis line style={white},
		ymax=16,
		ymin=0,
		xmin=-0.01,
		xmax=0.35,
		ytick={0,5,10,15},
		yticklabels={\textcolor{blue}{0},\textcolor{blue}{5},\textcolor{blue}{10},\textcolor{blue}{15}},
		ylabel = {\textcolor{blue}{$P_{MAE}$ (\%)}},
		legend style={at={(axis cs:0.125,14.88)},anchor=west, font=\scriptsize},
		legend style={draw=none},
		grid=major,
		grid style={dashed,gray!50}
		]	
		\addplot[style={fill=babyblue,mark=none}] coordinates {(0,0.0) (0.05,9.971596003152012) (0.1,9.594671387698176) (0.15,10.75339066248329) (0.2,11.979507176792406) (0.25,15.709627379853883) (0.3,13.173820445998894)};
		\addlegendentry{$P_{MAE}$}	
		\draw [thick,blue] (axis cs:0.35,0) -- node[]{} (axis cs:0.35,15); 
		\end{axis}
		
		\begin{axis}[
		width=5.5cm,		
		height=4cm,		
		axis x line*=bottom, 
		axis y line*=left,
		axis line style={white},
		ymax=1.4,
		ymin=0.98,
		xmin=-0.01,
		xmax=0.35,
		ytick={1.0,1.2,1.4},
		yticklabels={\textcolor{red}{1.0},\textcolor{red}{1.2},\textcolor{red}{1.4}},
		xtick={0,0.05,0.1,0.15,0.2,0.25,0.3},
		xticklabels={0,0.05,0.1,0.15,0.2,0.25,0.3},
		ylabel = {\textcolor{red}{Speedup}},
		xlabel = {Pruning rate},
		legend style={at={(axis cs:0,1.37)},anchor=west, font=\scriptsize},
		legend style={draw=none}]		
		\addplot [red, mark=square, thick] coordinates {(0,1) (0.05,1.202) (0.1,1.273) (0.15,1.277) (0.2,1.263) (0.25,1.281) (0.3,1.283)};
		\addlegendentry{Speedup}
		\draw [thick,red] (axis cs:-0.01,0.98) -- node[]{} (axis cs:-0.01,1.4);
		\draw [thick,black] (axis cs:-0.01,0.98) -- node[]{} (axis cs:0.35,0.98);
		\draw [draw=black] (axis cs: 0,1.34) rectangle (axis cs: 0.215, 1.4); 
		\end{axis}
		
		\end{tikzpicture}
	}\\
	\subfloat[]{
		\begin{tikzpicture}[scale=1]
		\pgfplotsset{
			scale only axis,
		}
		
		\begin{axis}[
		ybar,
		bar width=18pt,
		width=5.5cm,		
		height=4cm,		
		axis x line=none,
		axis y line*=right,
		axis line style={white},
		ymax=0,
		ymin=-6.5,
		xmin=-0.01,
		xmax=0.35,
		ytick={0,-2,-4,-6},
		yticklabels={\textcolor{blue}{0},\textcolor{blue}{-2},\textcolor{blue}{-4},\textcolor{blue}{-6}},
		ylabel = {\textcolor{blue}{$P_{MAE}$ (\%)}},
		legend style={at={(axis cs:0.135,-6)},anchor=west, font=\scriptsize},
		legend style={draw=none},
		grid=major,
		grid style={dashed,gray!50}
		]	
		\addplot[style={fill=babyblue,mark=none}] coordinates {(0,0.0) (0.05,-4.262351586708702) (0.1,-4.516575918695159) (0.15,-5.028650017467489) (0.2,-5.462172045218714) (0.25,-5.464248522302459) (0.3,-5.966273230391095)};
		\addlegendentry{$P_{MAE}$}	
		\draw [thick,blue] (axis cs:0.35,0) -- node[]{} (axis cs:0.35,-6.5); 
		\end{axis}
		
		\begin{axis}[
		width=5.5cm,		
		height=4cm,		
		axis x line*=bottom, 
		axis y line*=left,
		axis line style={white},
		ymax=1.42,
		ymin=0.98,
		xmin=-0.01,
		xmax=0.35,
		ytick={1.0,1.2,1.4},
		yticklabels={\textcolor{red}{1.0},\textcolor{red}{1.2},\textcolor{red}{1.4}},
		xtick={0,0.05,0.1,0.15,0.2,0.25,0.3},
		xticklabels={0,0.05,0.1,0.15,0.2,0.25,0.3},
		ylabel = {\textcolor{red}{Speedup}},
		xlabel = {Pruning rate},
		legend style={at={(axis cs:0.01,1.015)},anchor=west, font=\scriptsize},
		legend style={draw=none}]		
		\addplot [red, mark=square, thick] coordinates {(0,1) (0.05,1.388) (0.1,1.390) (0.15,1.408) (0.2,1.403) (0.25,1.411) (0.3,1.411)};
		\addlegendentry{Speedup}
		\draw [thick,red] (axis cs:-0.01,0.98) -- node[]{} (axis cs:-0.01,1.42);
		\draw [thick,black] (axis cs:-0.01,0.98) -- node[]{} (axis cs:0.35,0.98);
		\draw [draw=black] (axis cs: 0.01,0.985) rectangle (axis cs: 0.225, 1.045); 
		\end{axis}
		
		\end{tikzpicture}
	}
	\subfloat[]{
		\begin{tikzpicture}[scale=1]
		\pgfplotsset{
			scale only axis,
		}
		
		\begin{axis}[
		ybar,
		bar width=18pt,
		width=5.5cm,		
		height=4cm,		
		axis x line=none,
		axis y line*=right,
		axis line style={white},
		ymax=24,
		ymin=0,
		xmin=-0.01,
		xmax=0.35,
		ytick={0,10,20},
		yticklabels={\textcolor{blue}{0},\textcolor{blue}{10},\textcolor{blue}{20}},
		ylabel = {\textcolor{blue}{$P_{MAE}$ (\%)}},
		legend style={at={(axis cs:0.125,22.3)},anchor=west, font=\scriptsize},
		legend style={draw=none},
		grid=major,
		grid style={dashed,gray!50}
		]	
		\addplot[style={fill=babyblue,mark=none}] coordinates {(0,0.0) (0.05,11.991528124919824) (0.1,18.94636887585853) (0.15,20.080928159885065) (0.2,18.073061421109692) (0.25,19.633063474002103) (0.3,16.636404119776365)};
		\addlegendentry{$P_{MAE}$}	
		\draw [thick,blue] (axis cs:0.35,0) -- node[]{} (axis cs:0.35,24); 
		\end{axis}
		
		\begin{axis}[
		width=5.5cm,		
		height=4cm,		
		axis x line*=bottom, 
		axis y line*=left,
		axis line style={white},
		ymax=1.72,
		ymin=0.98,
		xmin=-0.01,
		xmax=0.35,
		ytick={1.0,1.2,1.4,1.6},
		yticklabels={\textcolor{red}{1.0},\textcolor{red}{1.2},\textcolor{red}{1.4},\textcolor{red}{1.6}},
		xtick={0,0.05,0.1,0.15,0.2,0.25,0.3},
		xticklabels={0,0.05,0.1,0.15,0.2,0.25,0.3},
		ylabel = {\textcolor{red}{Speedup}},
		xlabel = {Pruning rate},
		legend style={at={(axis cs:0,1.67)},anchor=west, font=\scriptsize},
		legend style={draw=none}]		
		\addplot [red, mark=square, thick] coordinates {(0,1) (0.05,1.576) (0.1,1.581) (0.15,1.589) (0.2,1.596) (0.25,1.600) (0.3,1.618)};
		\addlegendentry{Speedup}
		\draw [thick,red] (axis cs:-0.01,0.98) -- node[]{} (axis cs:-0.01,1.72);
		\draw [thick,black] (axis cs:-0.01,0.98) -- node[]{} (axis cs:0.35,0.98);
		\draw [draw=black] (axis cs: 0,1.62) rectangle (axis cs: 0.215, 1.72); 
		\end{axis}
		
		\end{tikzpicture}
	}
	\caption{Speedups and MAE when giving different pruning rate. (a) MovieLens 100K. (b) Appliances. (c) Book-Crossings. (d) Jester.}
	\label{fig:Speedup and MAE}
\end{figure}

We firstly set the number of latent factor dimensions, $k$, as 50, and measure the speedups and MAE of all datasets, in Figure \ref{fig:Speedup and MAE}.
As can be seen from Figure \ref{fig:Speedup and MAE}, the proposed methods realize 1.2-1.62 speedups, based on the given pruning rate on different dataset.
The runtime of the conventional training process is measured by setting the pruning rate as 0, so that no latent factors are eliminated.
Generally, by setting a greater pruning rate, our methods tend to provide a greater speedup, compared with the conventional training process, because more latent factors are pruned to reduce more computational time.
By definition, the time complexity of SVD algorithms, including FunkSVD, is $O(n^3)$ \cite{info11070369}. The time complexity of the MF algorithm accelerated by our proposed methods is also $O(n^3)$, but the total computational time can be theoretically reduced proportionally with the given pruning rate.
However, according to the experimental results, the speedup is not linearly increased when the pruning rate becomes larger.
This is because the proposed matrix rearrangement is based on the joint sparsity of the feature matrices, so some significant elements may have even larger indices than some insignificant ones, and thus are eliminated by the pruning process.
In addition, according to Figure \ref{fig:Distribution-T}, the feature matrices generally follow a normal distribution with its mean value close to 0, for different datasets and trained with different number of epochs. Thereby, a similar proportion of all latent factors would be considered as insignificant and are pruned, considering the relatively close threshold values determined by different pruning rates, so increasing the pruning rate may not significantly improve the speedup. Here we suggest to use the pruning rate as a parameter for our methods to trade off between speedup and prediction accuracy, rather than adjusting it to realize a certain speedup. 

According to Figure \ref{fig:Speedup and MAE}, greater errors are caused by the proposed methods, which can be expected because some latent factors are pruned, leading to approximate computation in the training process. Our approaches increase the MAE by up to 20.08\%.
However, one can always select different pruning rates to properly trade off between prediction accuracy and speedup.
It is mostly noticeable the overall error is even decreased by using our methods, for Book-Crossings shown in Figure \ref{fig:Speedup and MAE}(c).
It might be because $k$, given as 50, is too big to make the MF model fitting such a dataset.
Our approaches, pruning some of the latent factors, thereby cause better fitting in this case.
This result indicates our methods even help to prevent overfitting, when a large number of latent factor dimensions is inappropriately selected. 	

\begin{figure}[!htb]
	\centering
	\subfloat[]{
		\begin{tikzpicture}[font=\normalsize, scale=1]	
		\begin{axis}[
		width=7cm,		
		height=5cm,		
		axis x line*=bottom, 
		axis y line*=left,
		ymax=0.85,
		ymin=0.35,
		xmin=15,
		xmax=105,
		grid=major,
		grid style={dashed,gray!50},
		ylabel = {Runtime (second)},
		xlabel = {Number of latent vectors},
		xticklabels={20,40,60,80,100},
		xtick={20,40,60,80,100},
		name=main plot,
		legend style={at={(axis cs:21,0.78)},anchor=west},
		clip=false
		]
		\addplot coordinates {(20,3.9423e-01) (40,4.7796e-01) (60,6.0799e-01) (80,6.8339e-01) (100,8.1227e-01)};
		\addlegendentry{Conventional}
		\addplot coordinates {(20,3.7667e-01) (40,4.0242e-01) (60,4.1387e-01) (80,4.3044e-01) (100,4.5596e-01)};
		\addlegendentry{Accelerated}				
		\end{axis}
		\end{tikzpicture}
	}\hspace{0.3cm}
	\subfloat[]{
		\begin{tikzpicture}[font=\normalsize, scale=1]	
		\begin{axis}[
		width=7cm,		
		height=5cm,		
		axis x line*=bottom, 
		axis y line*=left,
		ymax=6,
		ymin=2.5,
		xmin=15,
		xmax=105,
		grid=major,
		grid style={dashed,gray!50},
		ylabel = {Runtime (second)},
		xlabel = {Number of latent vectors},
		xticklabels={20,40,60,80,100},
		xtick={20,40,60,80,100},
		name=main plot,
		legend style={at={(axis cs:21,5.5)},anchor=west},
		clip=false
		]
		\addplot coordinates {(20,2.9166e+00) (40,3.5605e+00) (60,4.3599e+00) (80,4.9361e+00) (100,5.6745e+00)};
		\addlegendentry{Conventional}
		\addplot coordinates {(20,2.8027e+00) (40,3.2207e+00) (60,3.3673e+00) (80,3.5906e+00) (100,3.7714e+00)};
		\addlegendentry{Accelerated}				
		\end{axis}
		\end{tikzpicture}
	}\\
	\subfloat[]{
		\begin{tikzpicture}[font=\normalsize, scale=1]	
		\begin{axis}[
		width=7cm,		
		height=5cm,		
		axis x line*=bottom, 
		axis y line*=left,
		ymax=11,
		ymin=4.5,
		xmin=15,
		xmax=105,
		grid=major,
		grid style={dashed,gray!50},
		ylabel = {Runtime (second)},
		xlabel = {Number of latent vectors},
		xticklabels={20,40,60,80,100},
		xtick={20,40,60,80,100},
		name=main plot,
		legend style={at={(axis cs:21,10)},anchor=west},
		clip=false
		]
		\addplot coordinates {(20,4.9866e+00) (40,6.3254e+00) (60,7.8867e+00) (80,9.2588e+00) (100,1.0748e+01)};
		\addlegendentry{Conventional}
		\addplot coordinates {(20,4.8910e+00) (40,5.4417e+00) (60,5.7083e+00) (80,5.9221e+00) (100,6.2737e+00)};
		\addlegendentry{Accelerated}				
		\end{axis}
		\end{tikzpicture}
	}\hspace{0.3cm}
	\subfloat[]{
		\begin{tikzpicture}[font=\normalsize, scale=1]	
		\begin{axis}[
		width=7cm,		
		height=5cm,		
		axis x line*=bottom, 
		axis y line*=left,
		ymax=30.5,
		ymin=12,
		xmin=15,
		xmax=105,
		grid=major,
		grid style={dashed,gray!50},
		ylabel = {Runtime (second)},
		xlabel = {Number of latent vectors},
		xticklabels={20,40,60,80,100},
		xtick={20,40,60,80,100},
		name=main plot,
		legend style={at={(axis cs:21,28)},anchor=west},
		clip=false
		]
		\addplot coordinates {(20,1.4551e+01) (40,1.8783e+01) (60,2.2771e+01) (80,2.6335e+01) (100,3.0331e+01)};
		\addlegendentry{Conventional}
		\addplot coordinates {(20,1.2705e+01) (40,1.3382e+01) (60,1.4088e+01) (80,1.4785e+01) (100,1.5756e+01)};
		\addlegendentry{Accelerated}				
		\end{axis}
		\end{tikzpicture}
	}
	\caption{Runtime when giving different number of latent factor dimensions. (a) MovieLens 100K. (b) Appliances. (c) Book-Crossings. (d) Jester.}
	\label{fig:Runtime}
\end{figure}
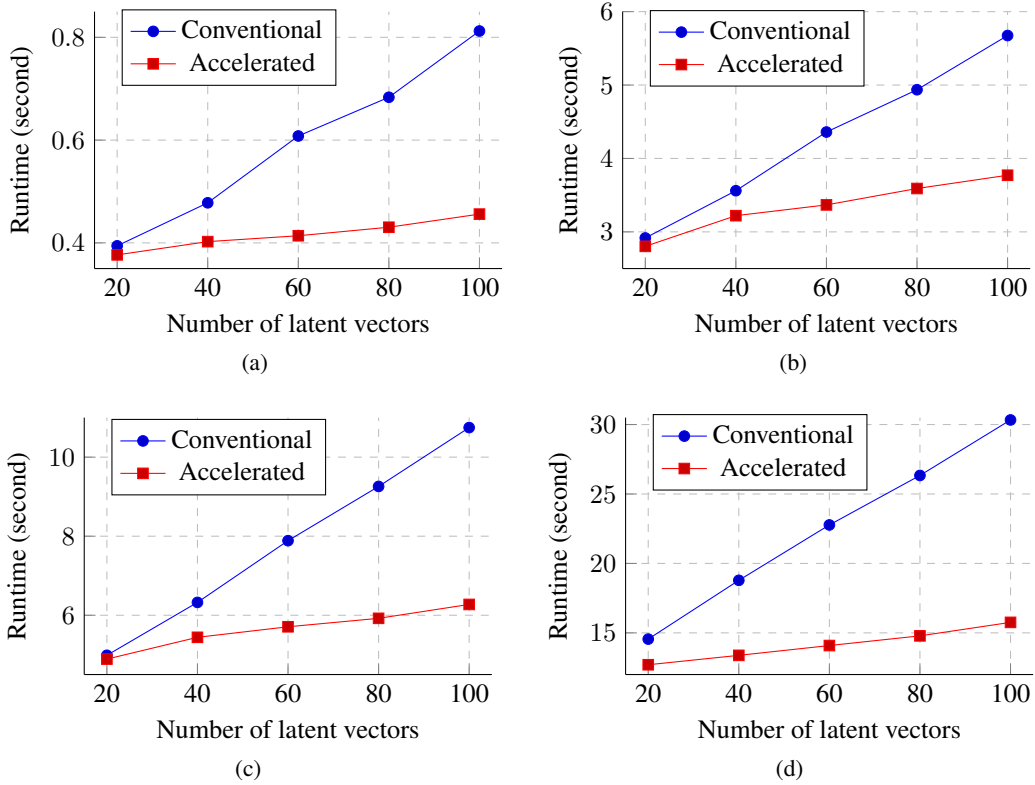

We then measure the runtime of the conventional process and the accelerated process using our methods, by setting different $k$, in Figure \ref{fig:Runtime}. Here we set the pruning rate as 0.3 for our methods.
As shown in Figure \ref{fig:Runtime}, the proposed methods only slightly reduce the overall computational time, when $k$ is small ($e.g.$, 20).
This is because the feature matrices are more dense in a small latent factor dimension.
An increased gap between the runtime of the conventional process and that of the accelerated process can be observed, suggesting a more dramatic speedup realized by the proposed approaches, when $k$ is increased, because more latent factors become insignificant and thus would be pruned by our methods.  

In practice, $k$ is often empirically determined. In this paper, we do not discuss how a suitable value of $k$ is determined.
However, as we have explained in Section \ref{sec:sparsity}, the fact that some users are less predictable and thus need a larger model for prediction, always remains.
A more steady increase of the runtime can be observed by using our methods, when $k$ is increased, so a RS can be designed with less considering the cost of longer computational time caused by larger $k$.

\subsection{Impact of Hyperparameters} \label{sec:impact}
Some hyperparameters of a MF-based RS might be selected according to the specific applications. 
Here we specifically investigate how a different learning rate, optimization strategy and initialization method can affect the speedup and prediction accuracy of our methods.

For the learning rate, we conduct experiments by setting the learning rates as 0.05, 0.1, 0.15, 0.2 and 0.25.
For the optimization strategy, we implement twin learners strategy \cite{r29} into LibMF.
The twin learners strategy allows some latent factors having a greater learning rate by not updating them in the first epoch and updating them normally afterwards, to alleviate the problem that the learning rate only dramatically changes in the first a few epochs, leading to local optimal.
For the initialization method, we further initialize all latent factors by a uniform distribution, instead of using a normal distribution, before the first training epoch.

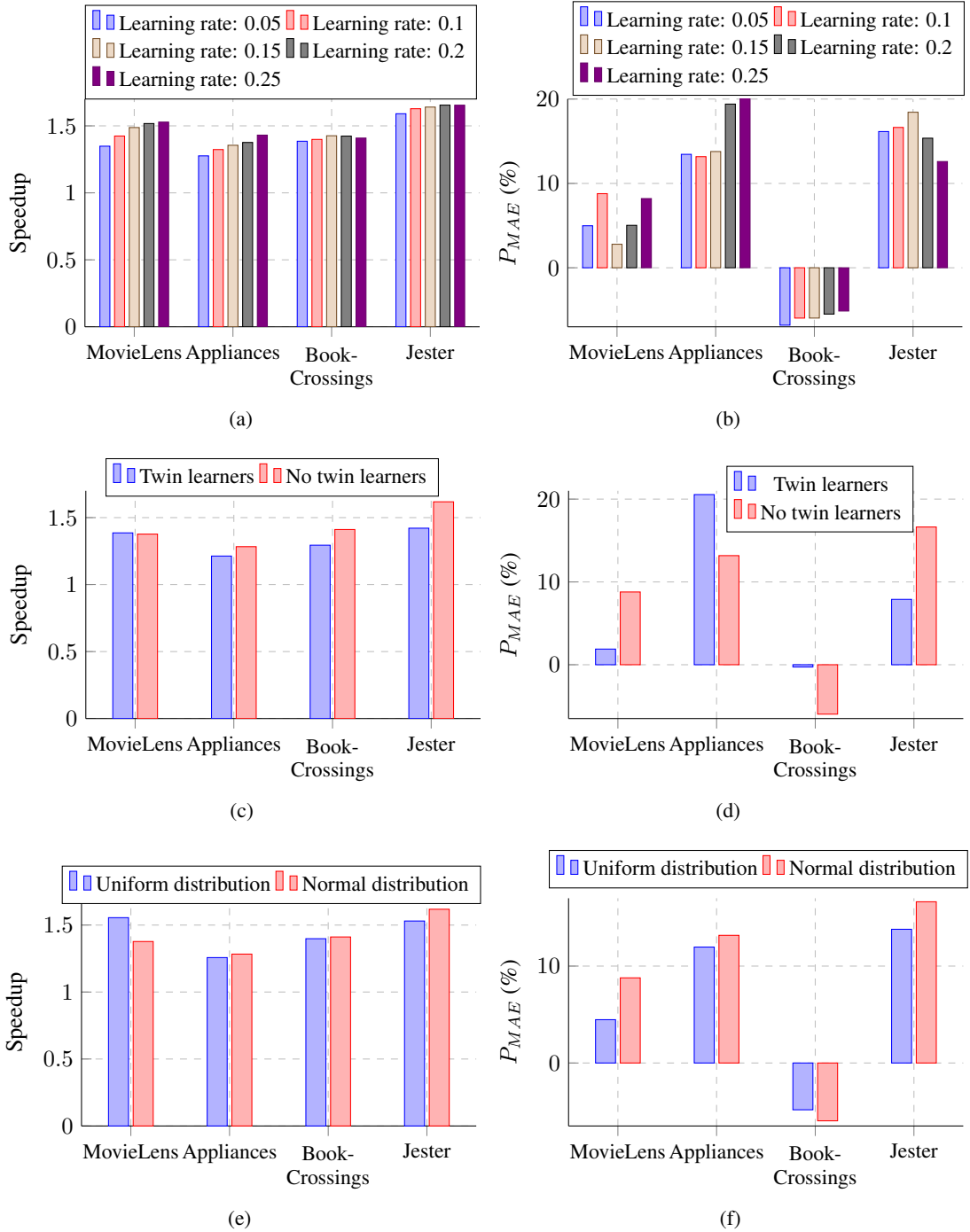
\begin{figure}[!htb]
	\centering
	\subfloat[]{
		\begin{tikzpicture}[font=\normalsize, scale=1]
		\begin{axis}[
		ybar,
		width=7.5cm,		
		height=5cm,		
		axis x line*=bottom, 
		axis y line*=left,
		ymax=1.7,
		ymin=0,
		xmin=-0.5, 
		xmax=3.5,
		xtick={0,1,...,3},
		xticklabels = {{MovieLens}, {Appliances}, {\begin{tabular}{c} Book- \\ Crossings \end{tabular}}, {Jester}},
		grid=major,
		grid style={dashed,gray!50},
		ylabel = {Speedup},
		ylabel near ticks,
		xlabel style={yshift=-0.4cm},
		bar width=0.15cm,
		xticklabel style = {font=\footnotesize},
		legend style={at={(axis cs:-0.5,2.05)},anchor=west,font=\footnotesize},
		legend columns=2
		]
		\addplot coordinates {(0,1.34820) (1,1.27610) (2,1.38517) (3,1.59048)};
		\addlegendentry{Learning rate: 0.05}
		\addplot coordinates {(0,1.42403) (1,1.32304) (2,1.39879) (3,1.62804)};
		\addlegendentry{Learning rate: 0.1}
		\addplot coordinates {(0,1.48742) (1,1.35505) (2,1.42602) (3,1.64027)};
		\addlegendentry{Learning rate: 0.15}
		\addplot coordinates {(0,1.51735) (1,1.37614) (2,1.42372) (3,1.65471)};
		\addlegendentry{Learning rate: 0.2}
		\addplot coordinates {(0,1.52812) (1,1.43083) (2,1.40880) (3,1.65428)};
		\addlegendentry{Learning rate: 0.25}
		\end{axis}
		\end{tikzpicture}
	}
	\subfloat[]{
		\begin{tikzpicture}[font=\normalsize, scale=1]
		\begin{axis}[
		ybar,
		width=7.5cm,		
		height=5cm,		
		axis x line*=bottom, 
		axis y line*=left,
		ymax=20,
		ymin=-7,
		xmin=-0.5, 
		xmax=3.5,
		xtick={0,1,...,3},
		xticklabels = {{MovieLens}, {Appliances}, {\begin{tabular}{c} Book- \\ Crossings \end{tabular}}, {Jester}},
		grid=major,
		grid style={dashed,gray!50},
		ylabel = {$P_{MAE}$ (\%)},
		ylabel near ticks,
		xlabel style={yshift=-0.4cm},
		bar width=0.15cm,
		xticklabel style = {font=\footnotesize},
		legend style={at={(axis cs:-0.45, 26)},anchor=west,font=\footnotesize},
		legend columns=2
		]
		\addplot coordinates {(0,4.981706782528914) (1,13.449006694968022) (2,-6.805658134663608) (3,16.14897321843941)};
		\addlegendentry{Learning rate: 0.05}
		\addplot coordinates {(0,8.785172328496746) (1,13.173820445998894) (2,-5.966273230391095) (3,16.636404119776365)};
		\addlegendentry{Learning rate: 0.1}
		\addplot coordinates {(0,2.7892908337504214) (1,13.779711494580674) (2,-5.963621484121006) (3,18.43938627052502)};
		\addlegendentry{Learning rate: 0.15}
		\addplot coordinates {(0,5.0313459521234964) (1,19.40351386532048) (2,-5.491635239680305) (3,15.36196286937061)};
		\addlegendentry{Learning rate: 0.2}
		\addplot coordinates {(0,8.203039992980992) (1,23.70035223670095) (2,-5.129173692703949) (3,12.593576401015923)};
		\addlegendentry{Learning rate: 0.25}
		\end{axis}
		\end{tikzpicture}
	}\\
	\subfloat[]{
		\begin{tikzpicture}[font=\normalsize, scale=1]
		\begin{axis}[
		ybar,
		width=7.5cm,		
		height=5cm,		
		axis x line*=bottom, 
		axis y line*=left,
		ymax=1.7,
		ymin=0,
		xmin=-0.5, 
		xmax=3.5,
		xtick={0,1,...,3},
		xticklabels = {{MovieLens}, {Appliances}, {\begin{tabular}{c} Book- \\ Crossings \end{tabular}}, {Jester}},
		grid=major,
		grid style={dashed,gray!50},
		ylabel = {Speedup},
		ylabel near ticks,
		xlabel style={yshift=-0.4cm},
		bar width=0.3cm,
		xticklabel style = {font=\footnotesize},
		legend style={at={(axis cs:-0.3,1.8)},anchor=west,font=\footnotesize},
		legend columns=2
		]
		\addplot coordinates {(0,1.386) (1,1.213) (2,1.294) (3,1.421)};
		\addlegendentry{Twin learners}
		\addplot coordinates {(0,1.377) (1,1.283) (2,1.411) (3,1.618)};
		\addlegendentry{No twin learners}
		\end{axis}
		\end{tikzpicture}
	}	
	\subfloat[]{
		\begin{tikzpicture}[font=\normalsize, scale=1]
		\begin{axis}[
		ybar,
		width=7.5cm,		
		height=5cm,		
		axis x line*=bottom, 
		axis y line*=left,
		ymax=21,
		ymin=-6.5,
		xmin=-0.5, 
		xmax=3.5,
		xtick={0,1,...,3},
		xticklabels = {{MovieLens}, {Appliances}, {\begin{tabular}{c} Book- \\ Crossings \end{tabular}}, {Jester}},
		grid=major,
		grid style={dashed,gray!50},
		ylabel = {$P_{MAE}$ (\%)},
		ylabel near ticks,
		xlabel style={yshift=-0.4cm},
		bar width=0.3cm,
		xticklabel style = {font=\footnotesize},
		legend style={at={(axis cs:1.1, 20)},anchor=west,font=\footnotesize},
		]
		\addplot coordinates {(0,1.889028346647911) (1,20.546174133914572) (2,-0.2733611350018092) (3,7.8897145711711065)};
		\addlegendentry{Twin learners}
		\addplot coordinates {(0,8.785172328496746) (1,13.173820445998894) (2,-5.966273230391095) (3,16.636404119776365)};
		\addlegendentry{No twin learners}
		\end{axis}
		\end{tikzpicture}
	}\\
	\subfloat[]{
		\begin{tikzpicture}[font=\normalsize, scale=1]
		\begin{axis}[
		ybar,
		width=7.5cm,		
		height=5cm,		
		axis x line*=bottom, 
		axis y line*=left,
		ymax=1.7,
		ymin=0,
		xmin=-0.5, 
		xmax=3.5,
		xtick={0,1,...,3},
		xticklabels = {{MovieLens}, {Appliances}, {\begin{tabular}{c} Book- \\ Crossings \end{tabular}}, {Jester}},
		grid=major,
		grid style={dashed,gray!50},
		ylabel = {Speedup},
		ylabel near ticks,
		xlabel style={yshift=-0.4cm},
		bar width=0.3cm,
		xticklabel style = {font=\footnotesize},
		legend style={at={(axis cs:-0.7,1.8)},anchor=west,font=\footnotesize},
		legend columns=2
		]
		\addplot coordinates {(0,1.555) (1,1.257) (2,1.398) (3,1.530)};
		\addlegendentry{Uniform distribution}
		\addplot coordinates {(0,1.377) (1,1.283) (2,1.411) (3,1.618)};
		\addlegendentry{Normal distribution}
		\end{axis}
		\end{tikzpicture}
	}
	\subfloat[]{
		\begin{tikzpicture}[font=\normalsize, scale=1]
		\begin{axis}[
		ybar,
		width=7.5cm,		
		height=5cm,		
		axis x line*=bottom, 
		axis y line*=left,
		ymax=17,
		ymin=-6.5,
		xmin=-0.5, 
		xmax=3.5,
		xtick={0,1,...,3},
		xticklabels = {{MovieLens}, {Appliances}, {\begin{tabular}{c} Book- \\ Crossings \end{tabular}}, {Jester}},
		grid=major,
		grid style={dashed,gray!50},
		ylabel = {$P_{MAE}$ (\%)},
		ylabel near ticks,
		xlabel style={yshift=-0.4cm},
		bar width=0.3cm,
		xticklabel style = {font=\footnotesize},
		legend style={at={(axis cs:-0.7, 20)},anchor=west,font=\footnotesize},
		legend columns=2
		]
		\addplot coordinates {(0,4.468335421877423) (1,11.96425646355315) (2,-4.828162743864415) (3,13.798368613700122)};
		\addlegendentry{Uniform distribution}
		\addplot coordinates {(0,8.785172328496746) (1,13.173820445998894) (2,-5.966273230391095) (3,16.636404119776365)};
		\addlegendentry{Normal distribution}
		\end{axis}
		
		\end{tikzpicture}
	}
	\caption{Speedups and $P_{MAE}$ when using our methods with different hyperparameters. (a) Speedups for different learning rates. (b) $P_{MAE}$ for different learning rates. (c) Speedups for different optimization strategies. (d) $P_{MAE}$ for different optimization strategies. (e) Speedups for different initialization methods.   (f) $P_{MAE}$ for different initialization methods.}
	\label{fig:hyperparameters}
\end{figure}

We demonstrate the speedups when using our methods with different hyperparameters, in Figures \ref{fig:hyperparameters} (a), (c) and (e), where the pruning rate is set to be 0.3.
According to Figure \ref{fig:hyperparameters} (a), a greater speedup can be realized by our methods when a larger learning rate is used. 
We provide the explanation as follows. 
According to Figure \ref{fig:Distribution-T}, the overall distribution of the latent factors becomes flatter with a larger deviation, as the number of epoch increases.
It can be expected that the overall distribution of the latent factors would also become flatter, when increasing the learning rate, as training process would be converged more quickly. 
Therefore, after the first epoch, our methods would lead to a larger threshold value, given the same pruning rate, when the learning rate is larger, causing more latent factors to be pruned in the subsequent iterations and resulting in a greater speedup.
According to Figure \ref{fig:hyperparameters} (a), a greater than 1.2 speedup can be realized by our methods in each case, indicating our methods are applicable when setting different learning rate.
Furthermore,
Since our methods can achieve high speedup at different learning rates, our methods are also applicable to other optimizers with adaptive learning rate like Adadelta and Adam.
According to Figures \ref{fig:hyperparameters} (c) and (e), the speedups realized by our methods are generally independent of the optimization strategies and initialization methods.

As shown in Figures \ref{fig:hyperparameters} (b), (d) and (f), the errors caused by our methods may vary when using different hyperparameters, but the $P_{MAE}$ is generally less than 20\% in each case. 
Furthermore, $P_{MAE}$ is mostly affected by dataset, instead of the hyperparameters, as the $P_{MAE}$ caused by different hyperparameters for a certain dataset is relatively similar, according to Figures \ref{fig:hyperparameters} (b), (d) and (f). For instance, greater $P_{MAE}$ can be observed for Appliances and Jester, compared with that for MovieLens and Book-Crossings, in all the cases considering certain learning rates, optimization strategies and initialization methods. 

The above results suggest the performance, such as speedup and error, might be affected when using our methods for different datasets, but the difference is mostly caused by the dataset itself, rather than the selection of certain hyperparameters, Thus, our methods are applicable considering different hyperparameters.

\section{Conclusions} \label{sec:conclusions}
This paper proposes some methods to accelerate the training process of MF-based RSs.
During the MF process, feature matrix multiplication and latent factor update are iteratively performed. However, we observe fine-grained structured sparsity in the decomposed feature matrices, which is inevitable because of the predictability for different users or items.
The fine-grained structured sparsity causes a large amount of unnecessary operations induced by irregularly located insignificant latent factors, increasing the computational time of both feature matrix multiplication and latent factor update.
To address this problem, we firstly propose to rearrange the feature matrices based on joint sparsity, to make a latent vector with a smaller index potentially more dense than that with a larger index in the feature matrices.
The feature matrix rearrangement is performed as pre-pruning process, to reduce the error caused by the later pruning process.
We further propose to dynamically prune the most insignificant latent factors from both matrix multiplication and gradient descent, during each epoch, to accelerate the training process.
The experimental results show our methods can realize 1.2-1.65 speedups, with up to 20.08\% error increase, compared with the conventional MF training process.
We show the speedup can be further increased if a larger latent factor dimension is applied in the RS.
In addition, we prove that our methods are applicable to be implemented in MF-based RSs considering different hyperparameters, including learning rate, optimization strategy and initialization method.
Finally, we highlight our methods, unlike all the existing techniques, which accelerate MF by putting in extra computational resources and parallel computing, can reduce the computational time of MF without any additional hardware resources.     

\appendix
\section{Appendix}
Here we provide the detailed derivation of Equations (7) and (8), which determine a threshold value according to a given pruning rate.
Based on the assumption that the original feature matrix follows normal distribution, we define $p$ as a given pruning rate, $F(x)$ as the cumulative distribution function of a feature matrix, $\mu$ and $\sigma$ as the mean and standard deviation from $F(x)$, respectively, and $\phi(x)$ as the cumulative distribution function of a standard Gaussian distribution.

The objective is to find a threshold $T$ ($T>0$) to meet the condition that the proportion of the elements located in the range between $-T$ and $T$ in $F(x)$ is $p$, as: 
\begin{gather}
F(T) - F(-T) = p \label{eq:app1}
\end{gather}

The value of $T$ that satisfies Equation (\ref{eq:app1}) can be determined, by firstly converting $F(x)$ into $\phi(x)$ and then searching the standard normal table. In such a case, we assume $-T$ and $T$ in $F(x)$ are relocated in $\phi(x)$ as, $x_1$ and $x_2$, respectively. According to the conversion from a Gaussion distribution to a standard one, we provide Equations (\ref{eq:app2}) and (\ref{eq:app3})
\begin{gather}
x_1 = (-T - \mu)/\sigma \label{eq:app2} \\
x_2 = (T - \mu)/\sigma \label{eq:app3}
\end{gather}
where $x_1$ and $x_2$ satisfy Equations (\ref{eq:app4}) and (\ref{eq:app5}) and Equation (\ref{eq:app5}) is derived from Equations (\ref{eq:app2}) and (\ref{eq:app3}).
\begin{gather}
\phi(x_2) - \phi(x_1) = p \label{eq:app4} \\
x_1 + x_2 = -2\mu / \sigma \label{eq:app5}
\end{gather}

According to Equations (\ref{eq:app4}) and (\ref{eq:app5}), $x_2$ would satisfy Equation (\ref{eq:app6}), so it can be easily found from the standard normal table. 
\begin{gather}
\phi(x_2) - \phi(-x_2 - 2\mu/\sigma) = p \label{eq:app6}
\end{gather}

Finally, once $x_2$ are determined, according to Equation (\ref{eq:app3}), $T$ is calculated as Equation (\ref{eq:app7}), as follows:
\begin{gather}
T = \sigma x_2 + \mu \label{eq:app7}
\end{gather}

\bibliographystyle{unsrt}
\bibliography{reference}  






\end{document}